\documentclass[aps,prd,amsmath,twocolumn,notitlepage,showpacs,superscriptaddress,nofootinbib,usenatbib]{revtex4-1}

\usepackage[normalem]{ulem}

\usepackage{epsfig,psfrag,graphics,verbatim}
\usepackage{graphicx}
\usepackage{dcolumn}
\usepackage{bm}
\usepackage{xcolor}
\usepackage{float}
\usepackage{adjustbox}
\usepackage{amsmath}
\usepackage{cleveref}

\begin{document}

\title{Testing Modified Gravity theory (MOG) with Type Ia Supernovae, Cosmic Chronometers and Baryon Acoustic Oscillations} 

\author{Carolina Negrelli}
\affiliation{Grupo de Astrof\'{\i}sica, Relatividad y Cosmolog\'{\i}a, 
        Facultad de Ciencias Astron\'{o}micas y Geof\'{\i}sicas, 
        Universidad Nacional de La Plata,
        Paseo del Bosque S/N 1900 La Plata, 
        Argentina}
\affiliation{CONICET, Godoy Cruz 2290, 1425 Ciudad Aut\'{o}noma de Buenos Aires, Argentina}

\author{Lucila Kraiselburd}
\affiliation{Grupo de Astrof\'{\i}sica, Relatividad y Cosmolog\'{\i}a, 
        Facultad de Ciencias Astron\'{o}micas y Geof\'{\i}sicas, 
        Universidad Nacional de La Plata,
        Paseo del Bosque S/N 1900 La Plata, 
        Argentina}
\affiliation{CONICET, Godoy Cruz 2290, 1425 Ciudad Aut\'{o}noma de Buenos Aires, Argentina}  

\author{Susana Landau}
\affiliation{Departamento de F\'{\i}sica and IFIBA, 
        Facultad de Ciencias Exactas y Naturales, 
        Universidad de Buenos Aires , 
        Ciudad Universitaria - Pab. I, 
        Buenos Aires 1428, 
        Argentina}
\affiliation{CONICET, Godoy Cruz 2290, 1425 Ciudad Aut\'{o}noma de Buenos Aires, Argentina}  

\author{Claudia G. Sc\'{o}ccola}
\affiliation{Grupo de Astrof\'{\i}sica, Relatividad y Cosmolog\'{\i}a, 
        Facultad de Ciencias Astron\'{o}micas y Geof\'{\i}sicas, 
        Universidad Nacional de La Plata,
        Paseo del Bosque S/N 1900 La Plata, 
        Argentina}
\affiliation{CONICET, Godoy Cruz 2290, 1425 Ciudad Aut\'{o}noma de Buenos Aires, Argentina}

\date{\today}

\begin{abstract}
 We analyse the MOdified Gravity (MOG) theory, proposed by Moffat, in a cosmological context.  We use data from Type Ia Supernovae (SNe Ia), Baryon Acoustic Oscillations (BAO) and Cosmic Chronometers (CC) to test  MOG predictions.  For this, we perform $\chi^2$ tests considering fixed values of $H_0$  and $V_G$, the self-interaction potential of one of the scalar fields in the theory. Our results show that the MOG theory is in agreement with all data sets for some particular values of $H_0$ and $V_G$, being the BAO data set the most powerful tool to test MOG predictions, due to its constraining power.
\end{abstract}

\maketitle


\section{Introduction}
\label{intro}
\par The failure of Newton's theory of Gravitation to successfully predict the motion of stars within galaxies and of galaxies in galaxy clusters was first reported by Oort~\cite{Oort1932} and Zwicky~\cite{Zwicky1937}. Later observations  of rotation curves in spiral galaxies~\cite{Rubin1978,Rubin1980} also pointed towards the same direction, namely: by assuming General Relativity (or its Newtonian counterpart) as the theory of gravity,  the visible matter cannot account for the shape of the rotation curves.   To solve this discrepancy, a component of non-luminous matter has been postulated. More recently,  observations of astrophysical objects over a large range of mass and spatial scales (such as, for example, galaxy clusters, spiral and dwarf galaxies)~\cite{1998ApJ...498L.107T,2013JCAP...07..008H,2018MNRAS.475..532O,2016A&A...594A..24P,2003A&A...397..899S,2008ApJ...684.1143X,2014ApJ...794...59K, 2015NatPh..11..245I} have also shown the need for including a {\it dark} component of matter if the assumption of General Relativity as the theory of gravity holds.

\par As a consequence, one of the main ingredients of the standard cosmological model ($\Lambda$CDM) is a component of matter that does not couple to the electromagnetic field but can be detected via its gravitational interactions. In this way,   this model is able to explain type Ia Supernovae data~\cite{2018ApJ...859..101S},  the Cosmic Microwave Background (CMB) power spectra~\cite{Planckcosmo2018}, data from Baryon Acoustic Oscillations~\cite{SDSS_DR7,BOSS,DES_Y1} as well as the formation of large scale structures~\cite{2013MNRAS.432..743N}. However, there is something missing in this picture, namely, the nature of this {\it dark} matter is currently unknown and none of the proposed candidates  has been detected yet in the laboratory, despite numerous dedicated experimental efforts~\cite{2019JPhG...46j3003S}. 

\par An alternative to explain the mismatch between the data and the predictions of the current theory of gravity relies on a modification of the latter. In this regard, a  first attempt was introduced by Milgrom in 1983~\cite{1983ApJ...270..365M}; the Modified Newtonian Dynamics (MOND) is a phenomenological proposal that follows from observations of galaxy rotation curves and the Tully-Fisher relation.  More recently, in 2004, Bekenstein proposed a relativistic version of this theory named TeVeS~\cite{2004PhRvD..70h3509B}. One of the main unsolved issues of this theory is that it is not able to explain simultaneously the rotation curve of galaxies and the strong lensing effect, as well as the observations of the Bullet cluster~\cite{Clowe2006}.  Finally, in 2006, Moffat formulated  the MOdified Gravity (MOG) theory in which  one massive vector field and three scalar fields are added  to the gravitational sector of the theory~\cite{2006JCAP...03..004M}. Moreover, this theory can predict successfully the motion within globular clusters, and clusters of galaxies~\cite{2008ApJ...680.1158M, 2014MNRAS.441.3724M, 2017EPJP..132..417M} as well as  the  rotation curves of spiral and dwarf galaxies~\cite{2013MNRAS.436.1439M,2017MNRAS.468.4048Z}. Nevertheless, some of us recently showed  that the MOG theory is not able to explain the observed rotation curve of the Milky Way~\cite{Negrelli2018}.
Moreover, when analyzing the Bullet cluster data~\cite{2006ApJ...648L.109C}, there are different claims in the literature: some authors affirm that  MOG cannot account for them~\cite{Clowe2006} while others hold that its predictions can  fit data from both the  Bullet and the Train Wreck merging clusters~\cite{2007MNRAS.382...29B,2016arXiv160609128I}. Additionally, it has been suggested that this theory is   not compatible with   the gas profile obtained from X-ray measurements and the strong-lensing properties of well-known galaxy clusters~\cite{2018MNRAS.tmp..370N}. However, in Ref   \cite{2017EPJP..132..417M} the authors show the opposite. On the other hand, as a consequence of  the detection of a neutron star merger followed by its electromagnetic counterpart by the LIGO experiment, theories where the difference between the velocity of gravitational waves and the speed of light  is significant, are ruled out~\cite{Boran2018}. However, Green et al.~\cite{Green2018} pointed out that this is the case for  bi-metric theories such as MOND and TeVeS, but not for MOG. 
In summary, the controversy about the compatibility of the MOG theory predictions with observational data at different scales is not yet settled.

\par On the other hand, it is well known  that type Ia supernovae data have led to the discovery of the current accelerated expansion of the Universe~\cite{Schmidt1998}. However, there is no agreement within the scientific community about the physical mechanism responsible for this phenomenon. The simplest candidate is the one adopted by the   $\Lambda$CDM model, namely, to include a cosmological constant in Einstein equations. Other proposals involve adding extra degrees of freedom to the Standard Model of Particles~\cite{Tsujikawa2011}. Alternatively, the extra degrees of freedom can be added to the gravitational sector of the theory.  In this regard, several alternative theories of gravity  have been considered~\cite{DeFelice2010}, being MOG a particular case  of such kind of theories. In summary, MOG offers an alternative to both the dark matter and dark energy ingredients of the standard cosmological model. 
\par In this work, we focus on the predictions of the MOG theory on cosmological scales. In order to test its predictions, we consider data from Supernovae type Ia, Baryon Acoustic Oscillations and Cosmic Chronometers. We use $\chi^2$ tests to carry out the comparison between the mentioned data sets and the theoretical predictions. For this, we consider two fixed values $H_0$ (the one inferred from local data and the one obtained from the CMB data) and several values of $V_G$, the self-interaction potential of the scalar field that represents the gravitational constant. 

\par In Section~\ref{theory}, we briefly characterize the main aspects of the MOG theory at cosmological scales, analyzing the Friedmann equations and the expressions of the scalar fields of the theory. In Section~\ref{method}, we describe in detail the data sets that are used to test the predictions of the theory. Results of the comparison between the predictions of the MOG theory and each data set considered in this paper  are shown and discussed in Section~\ref{results}. Finally we present our conclusions in Section~\ref{conclusions}.

\section{Cosmological background evolution of the MOG theory}
\label{theory}

\par In this section we summarize the main aspects of the MOG theory. In order to account for the effects that  in the standard cosmological model are produced by dark matter and dark energy,
this theory introduces  a massive vector field $\phi^\mu$ and three scalar fields which are the following: i) $G$, the gravitational coupling strength, which is promoted to a scalar field; ii) $\mu$, which corresponds to the mass of the vector field; and iii) $\omega$, which describes the coupling strength  between the vector field and matter. The gravitational and vector field $\phi^\mu$ actions are characterized by:
\begin{equation}
S_G=\frac{1}{16\pi}\int\frac{1}{G}\left({\it R}+2\Lambda\right)\sqrt{-g}~d^4x,
\end{equation}

\begin{eqnarray}
S_\phi&=&\frac{1}{4\pi}\int\omega\Big[\frac{1}{4}{\bf\it B^{\mu\nu}B_{\mu\nu}}-\frac{1}{2}\mu^2\phi_\mu\phi^\mu\nonumber\\
&+&V_\phi(\phi_\mu\phi^\mu)\Big]\sqrt{-g}~d^4x,
\end{eqnarray}
while the scalar fields action is given by:
\begin{eqnarray}
S_S=&\int\frac{1}{G}\Big[\frac{1}{2}g^{\alpha\beta}\biggl(\frac{\nabla_\alpha G\nabla_\beta G}{G^2}
+\frac{\nabla_\alpha\mu\nabla_\beta\mu}{\mu^2}+\nabla_\alpha \omega \nabla_\beta \omega \biggr)\nonumber\\
&+\frac{V_G(G)}{G^2}+\frac{V_\mu(\mu)}{\mu^2}+V_\omega(\omega)\Big]\sqrt{-g}~d^4x,
\label{scalar}
\end{eqnarray}
being $B_{\mu\nu}$ the Faraday tensor of the vector field ($B_{\mu\nu}=\partial_\mu\phi_\nu-\partial_\nu\phi_\mu$) and $V_\phi(\phi_\mu\phi^\mu)$, $V_G(G)$, $V_\omega(\omega)$ and $V_\mu(\mu)$,  the self-interaction potentials associated with the
vector field and the scalar fields. Finally, $\nabla_\nu$ is the covariant derivative with respect to the metric $g_{\mu\nu}$.

\par  The Friedmann equations for this theory can be obtained assuming an homogeneous and isotropic space-time which can be described by  Friedmann-Lema\^{i}tre-Robertson-Walker (FLRW) line element 
($ds^2 = dt^2 - a^2(t)[(1 - kr^2)^{-1}dr^2 + r^2 d\Omega^2])$~\cite{2009CQGra..26h5002M}:

\begin{eqnarray}\label{EqF1}
H^2+&\frac{k}{a^2}=\frac{8\pi G\rho}{3}-\frac{4\pi}{3}\biggl(\frac{\dot{G}^2}{G^2}+\frac{\dot{\mu}^2}{\mu^2}+\dot{\omega}^2- \frac{1}{4 \pi} G\omega\mu^2\phi_0^2\biggr)\nonumber\\
&+\frac{8\pi}{3}\biggl(\frac{\omega G V_{\phi}}{4 \pi}+\frac{V_{G}}{G^2}+\frac{V_{\mu}}{\mu^2}+V_{\omega}\biggr)+\frac{\Lambda}{3}+H\frac{\dot{G}}{G},
\end{eqnarray}
\begin{eqnarray}\label{EqF2}
\frac{\ddot{a}}{a}=&-\frac{4\pi G}{3}(\rho +3p)+\frac{8\pi}{3}\biggl(\frac{\dot{G}^2}{G^2}+\frac{\dot{\mu}^2}{\mu^2}+\dot{\omega}^2- \frac{1}{4 \pi}G\omega\mu^2\phi_0^2\biggr)\nonumber\\
&+\frac{8\pi}{3}\biggl(\frac{\omega G V_{\phi}}{4 \pi}+\frac{V_{G}}{G^2}+\frac{V_{\mu}}{\mu^2}+V_{\omega}\biggr)+\frac{\Lambda}{3}\nonumber\\
&+H\frac{\dot{G}}{2G}+\frac{\ddot{G}}{2G}-\frac{\dot{G}^2}{G^2},
\end{eqnarray}

these equations can be combined to obtain a differential equation for the scale factor $a$:\\

\begin{eqnarray}\label{juntas}
\frac{\ddot{a}}{a}+\frac{\dot{a}^2}{ 2 a^2}+\frac{k}{2 a^2}=&-4\pi G p+ 2 \pi\biggl(\frac{\dot{G}^2}{G^2}+\frac{\dot{\mu}^2}{\mu^2}+\dot{\omega}^2-\frac{G\omega\mu^2\phi_0^2}{ 4 \pi }\biggr)\nonumber\\
&+ 4\pi\biggl(\frac{\omega G V_{\phi}}{ 4 \pi }+\frac{V_{G}}{G^2}+\frac{V_{\mu}}{\mu^2}+V_{\omega}\biggr)\nonumber\\
&+\frac{\Lambda}{2}+\frac{\ddot{G}}{2 G}-\frac{\dot{G}^2}{G^2}+H\frac{\dot{G}}{2G}.
\end{eqnarray}

\par  In this work, like in the standard cosmological model, we model the baryonic matter  as a pressureless dust such that $p = 0$ and $w = p/\rho = 0$. On the other hand, it has been discussed in Section~\ref{intro} that in the MOG theory it is not necessary to include dark matter to explain  astrophysical and cosmological observations. Therefore, only baryonic matter will be included in the modified Friedmann Equations. However, given that the matter density  does not appear in Eq.~\ref{juntas}, the evolution of the cosmic scale factor and, in consequence, of the Hubble parameter  will depend only on the pressure of the baryonic matter which in the case of pressureless dust is zero.

\par On the other hand, the equation of motion for the scalar fields can be derived from Eq.~\ref{scalar}:

\begin{eqnarray}\label{EqF3}
\ddot{G}&+3H\dot{G}-\frac{3}{2}\frac{\dot{G}^2}{G}+\frac{G}{2}\biggl(\frac{\dot{\mu}^2}{\mu^2}+\dot{\omega}^2\biggr)-\frac{3}{G}V_G-V'_G\nonumber\\
&+G\Big[\frac{V_{\mu}}{\mu^2}+V_{\omega}\Big]+\frac{G}{8\pi}\Lambda-\frac{3G}{8\pi} \biggl(\frac{\ddot{a}}{a}+H^2 + \frac{k}{a^2}\biggr)=0, 
\end{eqnarray}
\begin{equation}\label{EqF4}
\ddot{\mu}+3H\dot{\mu}-\frac{\dot{\mu}^2}{\mu}-\frac{\dot{G}}{G}\dot{\mu}+\frac{G\omega \mu^3\phi^2_0}{ 4 \pi}+\frac{2}{\mu}V_{\mu}-V'_{\mu}=0,
\end{equation}
\begin{equation}\label{EqF5}
\ddot{\omega}+3H\dot{\omega}-\frac{\dot{G}}{G}\dot{\omega}+ \frac{1}{8 \pi} G\mu^2\phi^2_0 -\frac{GV_{\phi}}{ 4 \pi} - V'_{\omega}=0,
\end{equation}

where $V'_G =\frac{dV_G}{dG}$,  $V'_{\mu}=\frac{dV_\mu}{d\mu}$ and  $V'_{\omega}=\frac{dV_\omega}{d\omega}$.

\par In Section~\ref{intro}, we pointed out that the MOG theory might explain the late acceleration in the expansion of the Universe, even if  Einstein's cosmological constant is
set to 0. The reason for this is that the equation of state associated to   the self-interaction potentials $V_G$, $V_{\omega}$, $V_{\phi}$ and $V_{\mu}$  is of the form  $w_V = -1$~\cite{MT2007,2013Galax...1...65M}. One of the goals of this work is to test if the MOG theory can supply the role of the dark energy, therefore we set $\Lambda=0$. On the other hand, as  in  the standard cosmological model, we consider a flat universe, i.e.  $k=0$.

\par As regards the self-interaction potentials, the most 
simple assumption would be to take them all equal to 0. However, it has been shown that in order to describe properly cosmological quantities  such as the age and evolutionary stages of the Universe~\cite{2010arXiv1011.5174T,2016EPJC...76..490J}, and redshift space distortion (RSD) data~\cite{2018arXiv181104445J}, $V_G$ must be different from $0$. Furthermore, previous studies of the MOG theory in the cosmological context~\cite{MT2007,2013Galax...1...65M} consider $V_G >0$\footnote{The case $V_G>0$  has not  been analyzed in Refs.~\cite{2016EPJC...76..490J,2018arXiv181104445J}}.  Therefore, in this paper, we consider  MOG models with  $V_G=0$ and $V_G\neq0$ to test different data sets in the cosmological context (see section~\ref{results}). Also, we set $V_\mu=V_\omega=V_\phi=0$.

 \par To find the expressions for $H(z)$, $G(z)$, $\mu(z)$ and $\omega(z)$, equations~(\ref{juntas}),  (\ref{EqF3}), (\ref{EqF4}) and (\ref{EqF5})  need to be solved numerically with suitable initial conditions. We use the ones proposed in Ref.\cite{MT2007}:
\begin{eqnarray}
t_0&=&13.7\times 10^9~\mathrm{years},\nonumber\\
a_0&=&c \, t_0,\nonumber\\
G_0&=&6 \, G_N,\nonumber\\
\mu_0&=&a_0^{-1},\nonumber\\
\omega_0&=&1/\sqrt{12},\nonumber\\
\phi_0&=&0,\nonumber\\
\dot{a}_0&=&H_0 \, a_0,\nonumber\\
\dot{G}_0=\dot{\mu}_0=\dot{\omega}_0&=&0,\nonumber
\end{eqnarray}
where $G_N$ refers to the Newton's gravitational constant. The choices of  $t_0$ and  $a_0$, are the same as in the flat standard cosmological model. Besides, since there is no agreement within the scientific community about the reported value of $H_0$ (which is referred in the literature as the $H_0$ tension),  we consider two  different values  for this parameter, namely the values inferred from local and CMB observations  (see section~\ref{results} for a detailed discussion of this choice). The  solution of the MOG field equations in the spherically symmetric case (for details, see ~\cite{2009CQGra..26h5002M,2013Galax...1...65M}) defines the scalar fields values, being $\mu_0$ the inverse of the scale of the Universe. On the other hand, the choice of $G_0$ is motivated by the following reason:   an effective gravitational constant $G_{\rm eff}\approx 6 \, G_N $ at the Yukawa distance $r =\mu^{-1}$ is compatible with the results coming from the test particle equation of motion expression, 
being $G_{\rm eff}\approx 20 \, G_N $ at infinity. As a consequence, on superhorizon scales the present solution would be consistent with an Einstein-de Sitter cosmology without dark components~\cite{MT2007,2013Galax...1...65M}. 

\par
\begin{figure*}[t]
\centering
\includegraphics[width=0.9\textwidth, trim={0.5cm 0cm 0.5cm 0.5cm},clip=True]{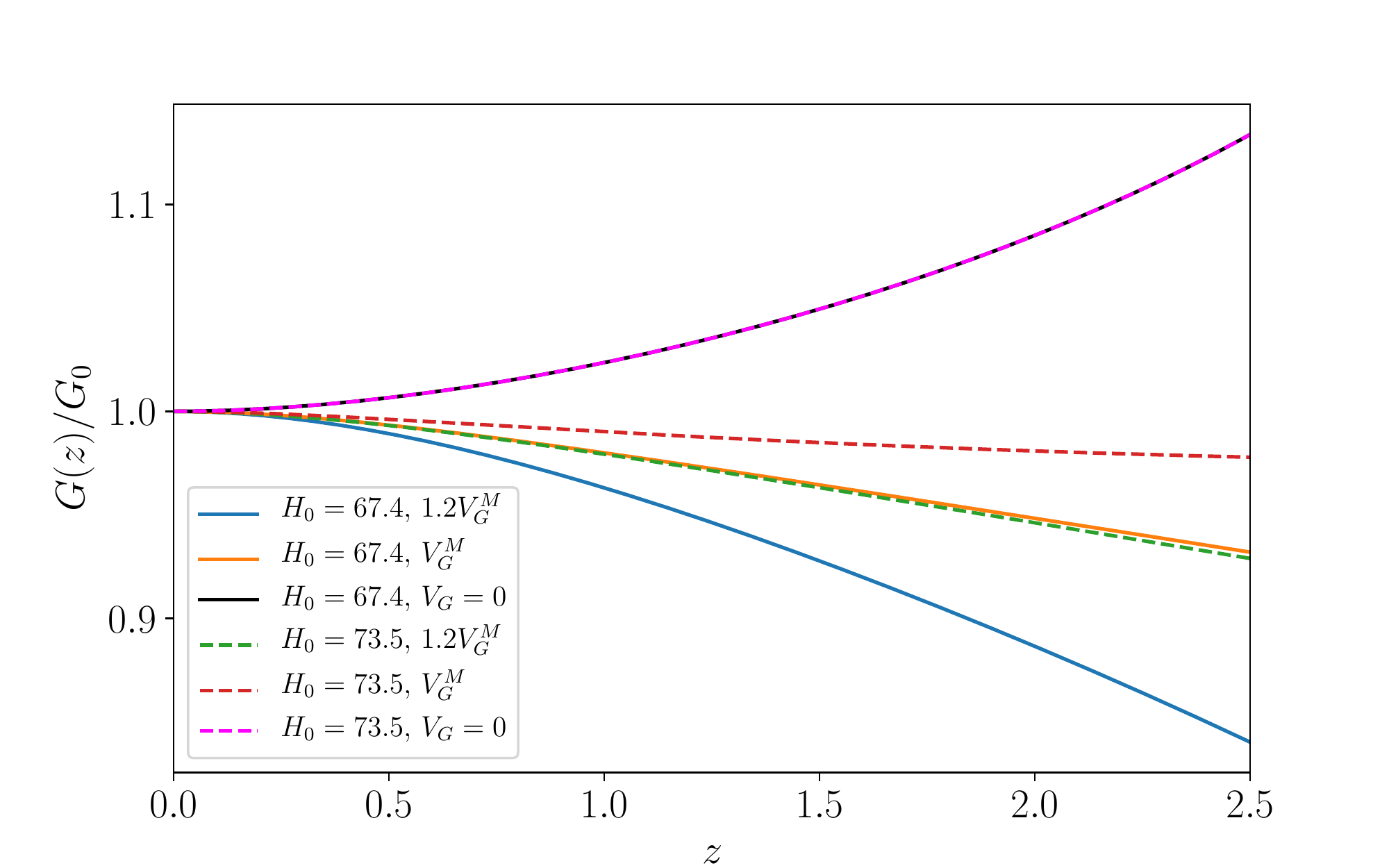}
\caption{$\frac{G(z)}{G_0}$ for  MOG models with fixed values of $H_0$ (labeled in the figure in units of {\rm km s$^{-1}$ Mpc$^{-1}$}) and $V_G$. $V_G^M = 0.077 \frac{G_0^2}{t_0^2}$ is the value suggested in Ref.~\cite{MT2007} which also analyses the MOG theory in the cosmological context.}
\label{fig:Hz}  
\end{figure*}

 The integrated $G(z)/G_0$ can be seen in Fig \ref{fig:Hz}. Also, as a result of the numerical integration, it follows  that $\omega$ and $\mu$ are constants as a function of $z$. All these results are consistent with the ones obtained by Moffat \& Toth in~\cite{MT2007}. On the other hand, it follows from Fig.~\ref{fig:Hz} that if $V_G=0$, the behaviour of   $G(z)/G_0$ does not depend on the value assumed for  $H_0$. Conversely,  if the case with $V_G\neq0$ is considered, different solutions are obtained when different values of $H_0$ are considered (see Fig.~\ref{fig:Hz}).  An explanation for this behavior lies in the fact that,  if $V_G\neq 0$, a term proportional to $\frac{V_G}{G^2}$ is added to Eq.~\ref{juntas}. 

\section{Observational constraints}
\label{method}

In this section, we describe the data sets that we use to make the comparison with the predictions of the MOG theory.

\subsection{Cosmic chronometers}

\par 

The Cosmic Chronometer (CC) approach is very useful to track the Universe evolution. This method was proposed and implemented in~\cite{2002ApJ...573...37J} and it  allows to determine the Hubble  parameter $H(z)$ from the following expression:

\begin{equation}
H(z)=\frac{-1}{1+z}\frac{dz}{dt}.
\end{equation}

The method is based on the differential age evolution  of old elliptical passive-evolving\footnote{Passive-evolving in the sense that there is no star formation or interaction with other galaxies.} galaxies  that formed at the same time but are separated by a small redshift interval. In this way, by measuring the age difference of those galaxies,  the derivative $\frac{dz}{dt}$  can be measured from the ratio $\frac{\Delta z}{\Delta t}$.

\par  Good candidates  for cosmic chronometers are   galaxies that have been formed early in the Universe, at high redshift ($z > 2-3$), with large mass ($\mathcal{M}_{\rm stars}>10^{11}\mathcal{M}_{\odot}$), and which have not had any star formation ever since. In this way, if these galaxies are observed at later cosmic time, the age evolution of their stars can be used as a clock synchronized   with the evolution of cosmic time. Spectroscopic surveys allow to obtain $dz$ with high accuracy. The main advantage of this method lies in   the measurement of relative ages $dt$, which avoids the systematic effects that affect the determination of absolute ages. Moreover, the determination of  $dt$ depends only on atomic physics and  not  on the integrated distance along the line of sight (redshift)
which, in turn, is a function of the cosmological model. 

\par In this paper, we consider the most recent and accurate estimates of $H(z)$ acquired through this method (see Table \ref{tab:CC}) to test the predictions of the MOG theory. Simon et al.~\cite{simon05} used the GDDS catalogue~\cite{Abraham2004} together with spectroscopic data from other galaxies~\cite{Dunlop1996,Nolan2003,Spinrad1997,Treu1999,Treu2001,Treu2002} to obtain 9 determinations of $H(z)$ in the redshift range $0.09 \leq z \leq 1.75$.  Stern et al.~\cite{stern10} obtained high quality spectra from red galaxies and used them together with spectra from the SPICES~\cite{Stern2001} and VVDS~\cite{LeFevre2005} catalogues to obtain two measurements of $H(z)$  at $z= 0.48$ and $z=0.88$. Moresco et al.~\cite{moresco12} took a sample of  11,000 massive red galaxies from  different catalogues~\cite{Cimatti2002,Demarco2010,Eisenstein2001,LeBorgne2006,Lilly2009,Onodera2010,Rosati2009,Stern2010,Strauss2002,Vanzella2008} to obtain 8 measurements of $H(z)$ within the redshift range $0.17 < z < 1.04$. Zhang et al.~\cite{zhang14}  obtained their results in the redshift interval $0.07 \leq z \leq 0.28$ taking 17,832 red galaxies from the Sloan Digital Sky Survey Data Release Seven (SDSS DR7)~\cite{Abazajian2009}.  Moresco~\cite{moresco15} considered spectroscopic data of 29 high redshift  galaxies  ($z >1.4$)~\cite{Gobat2013,Kriek2009,Krogager2014,Onodera2012,Saracco2005} to obtain estimations of $H(z)$ at redshift $z=1.363$ and $z=1.965$. Moresco et al.~\cite{CC2} used   data from the BOSS catalogue~\cite{Dawson2013,Eisenstein2011} to obtain 5 measurements of $H(z)$ in the redshift range $0.38 < z < 0.48$.

\bgroup
\def\arraystretch{1.3}
\begin{table}
\resizebox{0.3\textwidth}{!}{
\begin{tabular}{| c | c | c | c |}
    \hline
    $z$ & $H(z) \ (\mathrm{km}\,\mathrm{s}^{-1}\,\mathrm{Mpc}^{-1}) $ & Reference \\
    \hline
     0.09 & 69  $\pm$  12   & \\
    0.17 & 83  $\pm$  8    &  \\
    0.27 & 77  $\pm$  14   & \\
    0.4  & 95 $\pm$ 17     & \\
    0.9  &117 $\pm$ 23     &~\cite{simon05}  \\
    1.3  &168 $\pm$ 17     & \\
    1.43 &177 $\pm$ 18     & \\
    1.53 &140 $\pm$ 14     & \\ 
    1.75 &202 $\pm$ 40     & \\
    \hline
    0.48 & 97  $\pm$  62   &~\cite{stern10} \\
    0.88 & 90  $\pm$  40   & \\
    \hline  
    0.1791 &  75  $\pm$  4 & \\ 
    0.1993 & 75  $\pm$  5  & \\
    0.3519 & 83  $\pm$  14 & \\
    0.5929 & 104  $\pm$  13 &~\cite{moresco12} \\
    0.6797 & 92  $\pm$  8  & \\
    0.7812 & 105  $\pm$  12 & \\
    0.8754 & 125  $\pm$  17 & \\
    1.037  & 154  $\pm$  20 & \\
    \hline   
    0.07   & 69  $\pm$  19.6   &  \\
    0.12   & 68.6  $\pm$  26.2 &~\cite{zhang14} \\  
    0.2    & 72.9  $\pm$  29.6 & \\
    0.28   & 88.8  $\pm$  36.6 & \\ 
    \hline 
    1.363  & 160  $\pm$  33.6  &~\cite{moresco15}  \\
    1.965  & 186.5  $\pm$  50.4 & \\ 
    \hline 
    0.3802 & 83  $\pm$  13.5   &  \\
    0.4004 & 77  $\pm$  10.2   & \\
    0.4247 & 87.1  $\pm$  11.2 &~\cite{CC2}\\
    0.4497 & 92.8  $\pm$  12.9 & \\
    0.4783 & 80.9  $\pm$  9    & \\
\hline 
\end{tabular}}
\caption{$H(z)$ constraints obtained from the cosmic chronometers. The table shows the redshift of the measurement, the mean value and standard deviation of $H(z)$} in units of {\rm km s$^{-1}$ Mpc$^{-1}$  and the corresponding reference.}
\label{tab:CC}
\end{table}

\subsection{Supernovae type Ia}

\par Type Ia Supernovae (SNe Ia) are among the most energetic events in the Universe. They are quite common, and are observed in different kind of galaxies. The extremely  high luminosity of the supernovae makes them easily detectable by surveys. Most importantly, the homogeneity of their spectra and light curves makes them excellent candidates to be considered as standard candles.

The distance modulus of a supernova is given by,
\begin{equation}
\mu=25+5 \log_{10}(d_L(z)),
\label{distmod}
\end{equation}
being $d_L$ the luminosity distance that depends on the cosmological model and redshift $z$,
\begin{equation}
d_L(z)=(1+z)\int_0^z\frac{dz'}{H(z')}.
\label{distlum}
\end{equation}
Then, it is possible to compare the theoretical distance modulus of the MOG theory with the modulus obtained from the SNe Ia data. 

\par In this paper,  we consider  the Pantheon  compilation~\cite{2018ApJ...859..101S} of 1,048 SNe Ia in the redshift interval $0.01 < z < 2.3$. The observed distance modulus estimator for this compilation is given by
\begin{equation}
\mu=m_B-M+\alpha x_1+\beta c +\Delta_M+\Delta_B,
\label{mu}
\end{equation}
where $\Delta_M$ is a distance correction based on the mass of the SNe Ia's  host-galaxy, and $\Delta_B$ is a distance correction based on predicted biases from simulations. Furthermore, $m_B$ is an overall flux
normalization, $x_1$ and $c$ refer to the deviation from the average light-curve shape and the mean SNe Ia BV color, respectively\footnote{Parameters $m_B$, $x_1$ and $c$ result from the fit of a model of the SNe Ia spectral sequence to the photometric data, details in Ref.~\cite{2018ApJ...859..101S}.}. Finally, $M$ is the absolute B-band magnitude of a fiducial SNe Ia with $x_1 = 0$ and $c = 0$. Parameter $\alpha$ represents the coefficient of the relation between luminosity and stretch; while $\beta$, the coefficient of the relation between luminosity and color.\\
For the Pantheon data compilation  $\Delta_M$ is determined by:
\begin{equation}
\Delta_M=\gamma \times[1+e^{(-(m-m_{\rm step})/\tau)}]^{-1}
\end{equation}
where $\gamma$ stands for a relative offset in luminosity; $m_{\rm step}$, a mass step for the split; and $\tau$, an exponential transition term in a Fermi function that describes the relative probability of masses being on one side or the other of the split. The last two ($m_{\rm step}$ and $\tau$) are obtained from different host galaxies samples (technicalities are described in~\cite{2018ApJ...859..101S}) and $m$ is the host-galaxy mass. Finally, $\alpha$, $\beta$, $M$ and $\gamma$ account for nuisance parameters.

\subsection{BAO}
\label{bao}
Before the formation of neutral hydrogen, photons and free electrons were coupled through Thomson scattering, generating accoustic waves in the primordial plasma. After recombination, matter and radiation decouples.
The maximum distance the accoustic wave could travel defines a characteristic scale, named the sound horizon at the drag epoch $r_d$; this scale is imprinted in the distribution of matter in the Universe. Baryon Accoustic Oscillations (BAO) provide a standard ruler to measure cosmological distances. Different tracers of the underlying matter density field provide probes to measure distances at different redshifts.

Along the line of sight, the BAO signal directly constrains the Hubble constant $H(z)$ at redshift $z$. When measured in a redshift shell, it constrains the angular diameter distance $D_A(z)$,
\begin{equation}
    D_A(z) = 
    \frac{c}{(1+z)} \int_0^z \frac{dz'}{H(z')}.
\end{equation}
To separate
 $D_A(z)$ and $H(z)$, BAO should be measured in the 2D correlation function, for which extremely large volumes are necessary. If this is not the case, a combination of both quantities can be measured as

\begin{equation}
D_V(z) = \left[ (1+z)^2 D_A^2(z) \frac{cz}{H(z)}  \right]^{1/3}.
\end{equation}

BAO have been measured with great precision using different observational probes. To measure the BAO scale from the clustering of matter, it is necessary to define a fiducial cosmology. Most of the distance constraints presented in Table~\ref{tab:BAO_data} are multiplied by a factor $(r_d / r_d^{fid})$, which is the ratio between the sound horizon at  the drag epoch to the same quantity computed in the fiducial cosmology. We take this ratio as a free parameter in the statistical analysis.

\bgroup
\def\arraystretch{1.5}
\begin{table}
\resizebox{0.48\textwidth}{!}{
\begin{tabular}{| c | c | c | c |}
\hline 
 $z_{\rm eff}$ &  Value & Observable  & Reference  \\ 
 \hline 
$0.16$	& 	$456 \pm 27.0$ Mpc	&  $D_V$    &~\cite{6dFGS}\\
\hline
$0.15$	& 	$(664 \pm 25.0) (r_d / r_d^{fid})$ Mpc  &  $D_V$  &	\cite{SDSS_DR7}\\
\hline
$0.81$	& 	$(1649.5 \pm 66.0) (r_d /r_d^{fid})$  Mpc  &  $D_A$  &~\cite{DES_Y1}\\
\hline
$0.38$	& 	$(1512 \pm 33) (r_d /r_d^{fid})$  Mpc  &  $D_M$    & \\
$0.38$	& 	$ (81.2 \pm 3.2) (r_d /r_d^{fid})$  km s$^{-1}$ Mpc$^{-1}$ &  $H$   & \\
$0.51$	& 	$(1975 \pm 41) (r_d /r_d^{fid})$  Mpc  &  $D_M$   &~\cite{BOSS}\\
$0.51$	& 	$ (90.9 \pm 3.3) (r_d /r_d^{fid})$  km s$^{-1}$ Mpc$^{-1}$ &  $H$    & \\
$0.61$	& 	$(2307 \pm 50) (r_d /r_d^{fid})$  Mpc  &  $D_M$   & \\
$0.61$	& 	$ (99.0 \pm 3.4) (r_d /r_d^{fid})$  km s$^{-1}$ Mpc$^{-1}$ &  $H$    & \\
\hline
$0.44$	& 	$(1716 \pm 83.0) (r_d / r_d^{fid})$ Mpc  &  $D_V$  &	\\
$0.6$	& 	$(2221 \pm 101.0) (r_d / r_d^{fid})$ Mpc  &  $D_V$  &	\cite{WiggleZ}\\
$0.73$	& 	$(2516 \pm 86.0) (r_d / r_d^{fid})$ Mpc  &  $D_V$  &	\\
\hline
$1.52$	& 	$(3843 \pm 147.0) (r_d / r_d^{fid})$ Mpc  &  $D_V$  &	\cite{SDSS-IV_quasars}\\
\hline
$2.3$	& 	$(1336 \pm 45.7) (r_d / r_d^{fid})$ Mpc  &  $D_H$  &	\cite{SDSS-III_La_forests}\\
$2.3$	& 	$(5566 \pm 317.2) (r_d / r_d^{fid})$ Mpc  &  $D_M$  &	\\
\hline
$2.4$	& 	$(1327.4 \pm 53.0) (r_d / r_d^{fid})$ Mpc  &  $D_H$  &	\cite{La_forests_quasars_cross}\\
$2.4$	& 	$(5259.7 \pm 250.5) (r_d / r_d^{fid})$ Mpc  &  $D_M$  &	\\
\hline
\end{tabular}}
\caption{Distance constraints from BAO measurements of different observational probes. The table shows the redshift of the measurement, the mean value and standard deviation of the observable, the observable that is measured in each case and the corresponding reference.}
\label{tab:BAO_data}
\end{table}

 In the following, we describe the observations used in this work.  The large-scale correlation function of the 6dF Galaxy Survey (6dFGS)~\cite{6dFGS}, which is obtained from a K-band selected galaxy subsample with redshifts, determines a value for the isotropic angular diameter distance $D_V$ at effective redshift, $z_{\rm eff}$ of 0.106. 
The same quantity at $z_{\rm eff}= 0.15$ is computed in Ross et al.~\cite{SDSS_DR7}, using the main sample of SDSS-DR7 galaxies, with measured redshifts, in combination with a reconstruction method to alleviate the effect of non-linearities on the BAO scale.
The first year data release of the Dark Energy Survey~\cite{DES_Y1} provides a measurement of the angular diameter distance $D_A$ at $z_{\rm eff}=0.81$, using the projected two point correlation function of a sample of over 1.3 million galaxies with measured photometric redshifts, distributed over a footprint of 1336 deg$^2$. 
The final galaxy clustering data release of the Baryon Oscillation Spectroscopic Survey~\cite{BOSS}, provides measurements of the comoving angular diameter distance $D_M$ (related with the physical angular diameter distance by $D_M(z) = (1+z) D_A(z)$) and Hubble parameter $H$ from the BAO  method after applying a reconstruction method, for three partially overlapping redshift slices centred at effective redshifts 0.38, 0.51, and 0.61. 
Using the WiggleZ Dark Energy Survey~\cite{WiggleZ} and a reconstruction method, measurements of $D_V$ at effective redshifts of  0.44, 0.6, and 0.73 are provided.
With a sample of 147000 quasars from the extended Baryon Oscillation Spectroscopic Survey (eBOSS)~\cite{SDSS-IV_quasars} distributed over 2044 square degrees with redshifts $0.8 < z < 2.2$, a measurement of $D_V$ at $z_{\rm eff}= 1.52$ is provided.
The BAO can be also determined from the flux-transmission correlations in Ly$\alpha$ forests in the spectra of 157,783 quasars in the redshift range $2.1 < z < 3.5$ from the Sloan Digital Sky Survey (SDSS) data release 12 (DR12)~\cite{SDSS-III_La_forests}. Measurements of $D_M$ and the Hubble distance (defined as $D_H = c/H(z)$) at $z_{\rm eff}=2.33$ are provided.
From the cross-correlation of quasars with the Ly$\alpha$-forest flux transmission of the final data release of the SDSS-III~\cite{La_forests_quasars_cross}, a measurement of $D_M$ and $D_H$ at $z_{\rm eff}=2.4$ can be obtained.


\section{Results}
\label{results}
\par  In this section, we compare  MOG cosmological predictions, explained in Section~\ref{theory}, with the observational data described in Section~\ref{method}. We use $\chi^2$ tests to quantify the agreement between the theoretical results and the data. 
To proceed with the comparison, we
define a fiducial model,  that  will be taken as  a reference to analyze the predictions of the MOG models.  It is well known that there is a tension  between the value of $H_0$ obtained from the Cosmic Microwave Background (CMB) ~\cite{Planckcosmo2018} and the one inferred  from local measurements ~\cite{Riess2018}.  Therefore, we choose two $\Lambda$CDM models with fixed values of $H_0$  and the total matter density parameter $\Omega_m$ as follows:
\begin{itemize}
    \item  $\Lambda$CDM Model 1: $H_0 =67.4$ {\rm km sec$^{-1}$ Mpc$^{-1}$ }and $\Omega_m=0.315$; the values obtained by the Planck collaboration using CMB data~\cite{Planckcosmo2018}.
    \item $\Lambda$CDM Model 2: $H_0 = 73.5$ {\rm km sec$^{-1}$ Mpc$^{-1}$ } and $\Omega_m = 0.298$; the values inferred  from cepheids and type Ia supernovae in Refs.~\cite{Riess2018} and~\cite{2018ApJ...859..101S} respectively.
\end{itemize}

 We perform separate statistical analyses for each data set, for which we test the  predictions of the MOG theory  for different values of $V_G$ and particular values of $H_0$ (namely, those of the fiducial models defined above).
 Results are shown in Tables \ref{tab:chi2_snia}, \ref{tab:chi2_snia_free}, \ref{tab:chi2_cc} and \ref{tab:chi2_BAO} together with the 
 corresponding $\tilde{\chi}^2_{5\sigma}$ value \footnote{If we assume that the probability distribution function  of the reduced $\chi^2$  for  a given number of data and free parameters is gaussian, its  mean value (which corresponds to the maximum probability) is equal to 1, and a $\sigma$ value for the dispersion can be defined. Furthermore, the 99.99995 \%  of probability is asigned to the confidence interval $1 \pm (\tilde{\chi}^2_{5\sigma}$ - 1), i.e. the confidence interval at $5\sigma$.}, for the number of data and free parameters considered in each case. Notice  that  $V_G^M=0.077 G_0^2/t_0^2$  is the value suggested by~\cite{MT2007}, who also analyze this theory in  a cosmological context.

It follows from Section~\ref{theory} that the theoretical prediction of the MOG theory for the scale factor evolution and its derived quantity $H(z)$ involves no free parameters. On the other hand, eq.~(\ref{mu}) shows that the analysis of supernovae data involves several free parameters: the nuisance parameters and the absolute magnitude $M$. In all  of the analyses done in this paper, we consider the absolute magnitude $M$  as a free parameter. Regarding the  nuisance parameters, we  consider two cases: i) fixed values given by the Pantheon compilation (Table \ref{tab:chi2_snia}) and ii) the nuisance parameters are allowed to vary (Table \ref{tab:chi2_snia_free}). The reason for this, is that the nuisance parameters given by the Pantheon compilation were obtained assuming a $\Lambda$CDM model for the theoretical predictions of the distance modulus and therefore it is not correct to assume {\it a priori}   those values  when analyzing an alternative cosmological model.  Nevertheless,  the obtained confidence intervals for the nuisance parameters are consistent with those given by the Pantheon compilation at 1$\sigma$ level. Therefore, the present analysis confirms the robustness of those parameters. 

It can be noticed from Table \ref{tab:chi2_snia} that only when $V_G=0$ is considered, the predictions of the corresponding MOG models are inconsistent with type Ia supernovae data  at  5$\sigma$. Conversely, the corresponding predictions of  all other MOG models considered in this paper, show agreement with the data within 3-$\sigma$. Accordingly,  Fig \ref{fig:snia} shows that there is a tiny difference between  the theoretical predictions for a  MOG model  with $V_G=V_G^M$, $H_0=63.4$ {\rm km sec$^{-1}$ Mpc$^{-1}$ } and the respective of $\Lambda$CDM model 1, while the model with $V_G =0$  fails to predict the behavior of the data.
Furthermore, Fig. \ref{fig:snia_2} shows that the difference  between the predictions of the $\Lambda$CDM model 2 and the MOG model  with $V_G=V_G^M$ and $H_0=73.5$ {\rm km sec$^{-1}$ Mpc$^{-1}$ }, while still small, is greater than the difference between the $\Lambda$CDM model 1 and the MOG model with  $V_G=V_G^M$ and $H_0=63.4$ {\rm km sec$^{-1}$ Mpc$^{-1}$}.   This might indicate  that the MOG theory could be a candidate to alleviate the $H_0$ tension.

Similar to the analysis with type Ia supernovae, when cosmic chronometers are considered, results in Table \ref{tab:chi2_cc} show that there is no agreement between the theoretical predictions of MOG models with $V_G=0$ and the data within 5-$\sigma$. Furthermore, not all MOG models with $V_G \neq 0$ and $H_0=73.5$ {\rm km sec$^{-1}$ Mpc$^{-1}$ } are in agreement with cosmic chronometers data within 5$\sigma$; which is the case  if $H_0=63.4$ {\rm km sec$^{-1}$ Mpc$^{-1}$} is assumed. On the other hand,   Figure~\ref{fig:cc} shows that the predictions for $H(z)$ change with the selected value of $H_0$. 

Finally, Table~\ref{tab:chi2_BAO}  shows that only two MOG models with $H_0=67.4$ {\rm km sec$^{-1}$ Mpc$^{-1}$ } have theoretical predictions that are in agreement with BAO data  within 2$\sigma$ while for the case $H_0=73.5$ {\rm km sec$^{-1}$ Mpc$^{-1}$ } only one model can explain the data within 5$\sigma$. Besides, all  other cases considered in this paper show a reduced $\chi^2$ value  beyond  the customary 5-$\sigma$ equivalent $\tilde{\chi}^2_{5\sigma}$. This behavior can also be appreciated in Figs. \ref{fig:bao} and \ref{fig:bao2}. Therefore, it should be noted that BAO data, which comprise several independent data sets, as described in Section~\ref{bao}, provide a useful tool to validate the predictions of the different MOG models analyzed in this paper. On the contrary, the predictions of the MOG models for type Ia supernovae data are very similar to the $\Lambda$CDM fiducial model's ones, provided $V_G\neq 0$.  Regarding  the cosmic chronometers, even though the predictions for the MOG models do not match the $\Lambda$CDM fiducial model's one, the large error bars in Fig.~\ref{fig:cc} prevent this data set to be more conclusive  when testing the MOG models.

\bgroup
\def\arraystretch{1.5}
\begin{table}
\resizebox{0.48\textwidth}{!}{
\begin{tabular}{|c|c|c|c|c|}
\hline
$V_G$ & 
\multicolumn{2}{c|}{$M$}&    \multicolumn{2}{c|}{$\tilde{\chi}^2$} \\
\cline{2-3} \cline{4-5}
    & $H_0=67.4$ & $H_0=73.5$ & $H_0=67.4$ & $H_0=73.5$ \\
\hline
$0$ & $-19.20 \pm 0.0042$ & $-19.01 \pm 0.0042$ & 1.40 & 1.40 \\
$0.6V_G^M$ & $-19.33 \pm 0.0042$ & $-19.12 \pm 0.0042$ &  1.08 & 1.12\\
$0.8V_G^M$ & $-19.38 \pm 0.0042$ & $-19.16 \pm 0.0042$ & 1.01 & 1.05 \\
$0.9V_G^M$ & $-19.40 \pm 0.0042$ & $-19.18 \pm 0.0042$ & 0.99 & 1.02\\
$V_G^M$ & $-19.43 \pm 0.0042$ & $-19.20 \pm 0.0042$ & 0.98 & 1.00\\
$1.1V_G^M$ & $-19.46 \pm 0.0042$ & $-19.22 \pm 0.0042$ & 0.99 & 0.98 \\
$1.2V_G^M$ & $-19.49 \pm 0.0042$ & $-19.25 \pm 0.0042$ & 1.03 &  0.98\\
\hline
\end{tabular}}
\caption{Results of the statistical analysis performed with Supernovae type Ia data with fixed nuisance parameters. The table shows the fixed values of $H_0$ and $V_G$ considered in each case, the obtained best fit value and 1-$\sigma$ error for the absolute magnitude $M$ and the corresponding reduced $\chi^2$-value. The equivalent $\chi^2$ at 5$\sigma$ is  $\tilde{\chi}^2_{5\sigma}=1.23$, while for  $\Lambda$CDM fiducial models 1 and 2 we obtain  $\tilde{\chi}^2_{\Lambda{\rm CDM}}=0.98$. $V_G^M$ is the value suggested in Ref.~\cite{MT2007} which also analyses the MOG theory in the cosmological context.}
\label{tab:chi2_snia}
\end{table}

\begin{figure*}[t]
\centering
\includegraphics[width=0.9\textwidth, trim={0.8cm 6.5cm 2.5cm 7cm},clip=True]{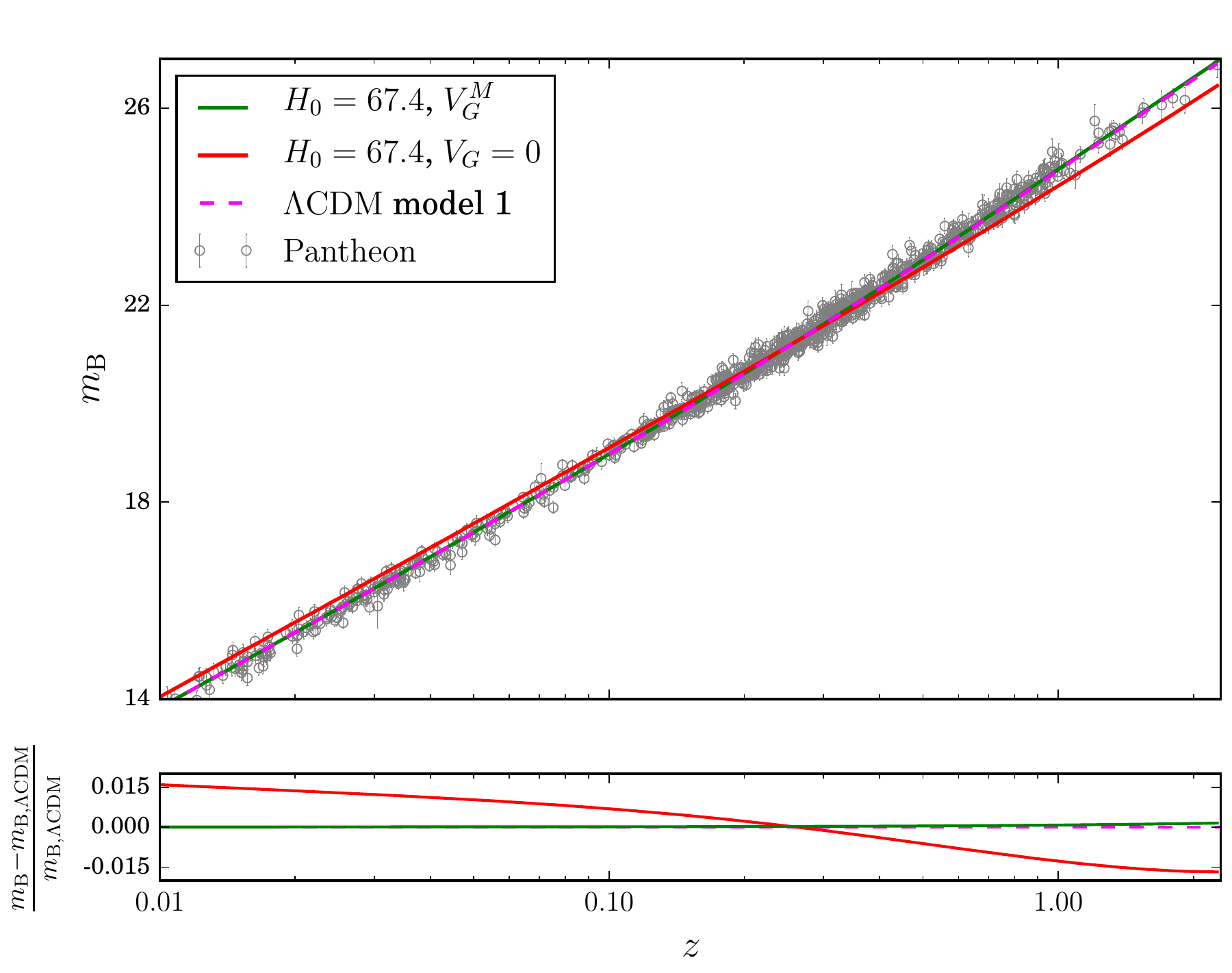}
\caption{Upper panel: $m_{\rm B}$ for the fiducial $\Lambda$CDM model 1  and  MOG models considering $H_0=67.4$ km sec$^{-1}$ Mpc$^{-1}$  and fixed  values of  $V_G$.  The data of the Pantheon compilation are shown in gray circles. The  $\Lambda$CDM fiducial model 1 and the MOG model with  $V_G=V_G^M$ are in agreement with the data within  5$\sigma$ while for the MOG model with   $V_G=0$ the obtained reduced $\chi^2$  value falls beyond the  5-$\sigma$ equivalent. Lower panel: relative differences between the $\Lambda$CDM model 1 and the MOG theory for different values for $H_0$ and $V_G$.}
\label{fig:snia}  
\end{figure*}

\begin{figure*}[t]
\centering
\includegraphics[width=0.9\textwidth, trim={0.8cm 6.5cm 2.5cm 7cm},clip=True]{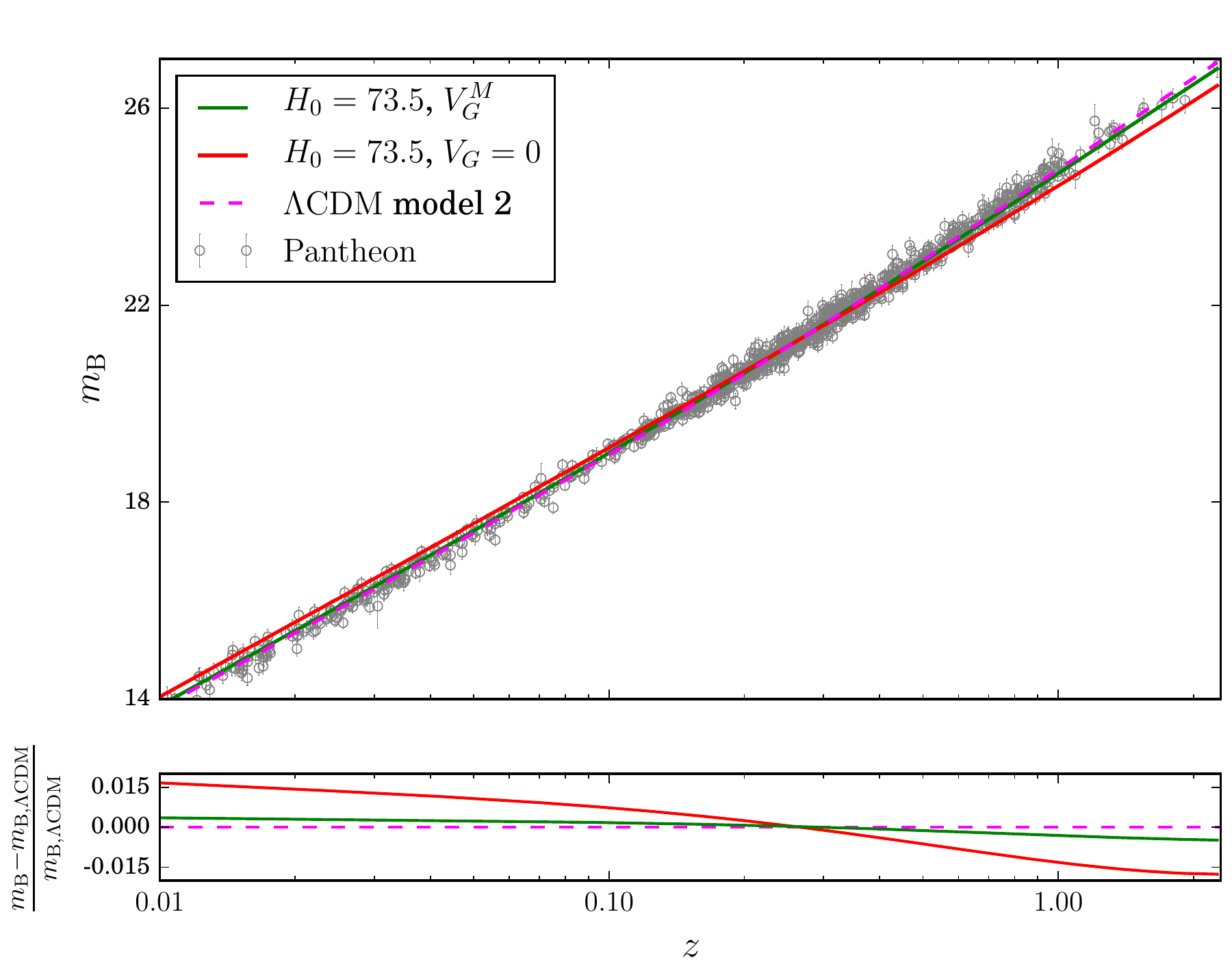}
\caption{Upper panel: $m_{\rm B}$ for the fiducial $\Lambda$CDM model 2 and  MOG models considering $H_0=73.5$ km sec$^{-1}$ Mpc$^{-1}$ and fixed  values of  $V_G$.  The data of the Pantheon compilation are shown in gray circles. The  $\Lambda$CDM fiducial model 2 and the MOG model with  $V_G=V_G^M$ are in agreement with the data within  5$\sigma$ while for the MOG model with $V_G=0$ the obtained reduced $\chi^2$  value falls beyond the  5-$\sigma$ equivalent. Lower panel: relative differences between $\Lambda$CDM model 2 and the MOG theory for different values for $H_0$ and $V_G$.}
\label{fig:snia_2}  
\end{figure*}


\bgroup
\def\arraystretch{1.6}
\begin{table}
\resizebox{0.5\textwidth}{!}{
\begin{tabular}{| c | c | c | c | c | c |}
\hline 
 $H_0$ &  $\alpha$ & $\beta$ & $\gamma$ & $M$ & $\tilde{\chi}^2$ \\ 
 \hline 
$67.4$	& 	-	&-&	-&	$-19.43\pm 0.004$	& $0.98$ \\
$73.5$	&-&	-&	-&	$-19.20\pm 0.004$ & $1.00$ \\
$67.4$	& $0.158\pm 0.005$ &	$3.03\pm  0.05$ &	$0.051\pm 0.009$ &	$-19.43\pm 0.007$ & $0.99$ \\
$73.5$ & $0.156\pm 0.005$ & $3.00\pm 0.05$	& $0.056  \pm 0.009$	& $-19.19\pm 0.007$ & $1.01$ \\
\hline
Pantheon & $0.154\pm 0.006$ & $3.02\pm 0.06$	& $0.053  \pm 0.009$	& &  \\
 \hline
\end{tabular}}
\caption{Results of the statistical analysis performed with Supernovae type Ia data. In all cases $V_G=V_G^M$. The first column shows the fixed values of $H_0$  considered in each case.  The first two entries  show the results for fixed nuisance parameters given by the Pantheon compilation  (also shown in Table \ref{tab:chi2_snia}).  The third and fourth entries show the results for the case where the nuisance parameters are allowed to vary. The equivalent $\chi^2$ at 5$\sigma$ for the latter is  $\tilde{\chi}^2_{5\sigma}=1.23$. The last entry shows the values of the nuisance parameters obtained by the Pantheon compilation.}
\label{tab:chi2_snia_free}
\end{table}

\bgroup
\def\arraystretch{1.4}
\begin{table}
\begin{tabular}{|c|c|c|}
\hline
$V_G$ & 
    \multicolumn{2}{c|}{$\tilde{\chi}^2$} \\
\cline{2-3}
    & $H_0=67.4$ & $H_0=73.5$\\
\hline
$0$ & 7.87 & 13.64 \\
$0.6V_G^M$ & 1.89 & 5.57\\
$0.8V_G^M$ & 0.78 & 3.41 \\
$0.9V_G^M$ & 0.52 & 2.49\\
$V_G^M$ & 0.53 & 1.71\\
$1.1V_G^M$ & 0.94 & 1.10 \\
$1.2V_G^M$ & 1.94 &  0.71\\
\hline
\end{tabular}
\caption{Results of the statistical analysis performed with cosmic chronometers. The table shows the fixed values of $H_0$ and $V_g$ considered in each case,  and the reduced $\chi^2$-value. The equivalent $\chi^2$ at 5$\sigma$ is  $\tilde{\chi}^2_{5\sigma}=2.79$ while for the $\Lambda$CDM fiducial models we obtain:  $\tilde{\chi}^2_{\Lambda{\rm CDM1}}=0.5$ and $\tilde{\chi}^2_{\Lambda{\rm CDM2}}=0.74$.}
\label{tab:chi2_cc}
\end{table}

\begin{figure*}[t]
\centering
\includegraphics[width=0.9\textwidth, trim={0.7cm 0cm 1.2cm 0cm},clip=True]{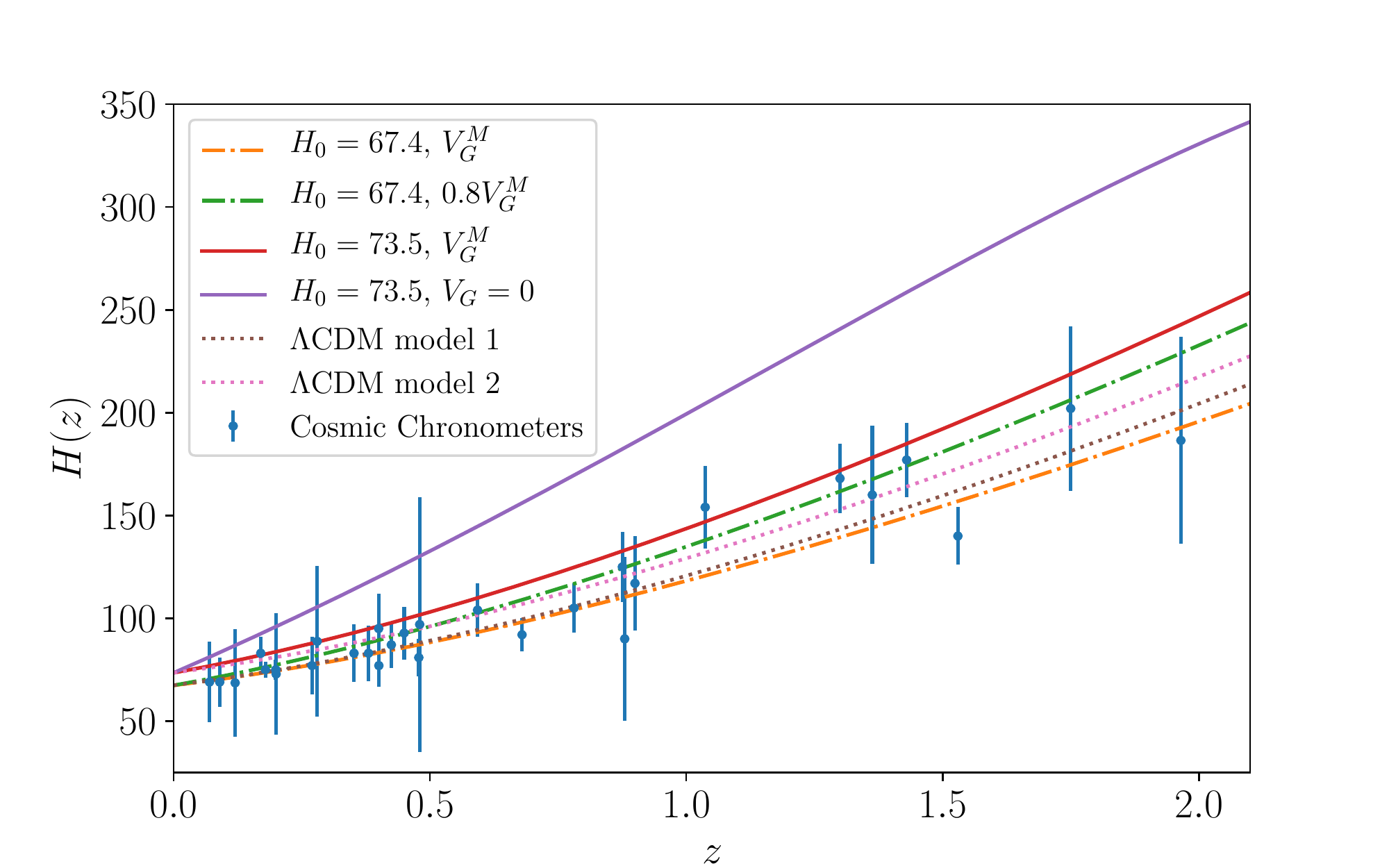}
\caption{$H(z)$ for the $\Lambda$CDM fiducial models 1 and 2 and the MOG models considering different values for $H_0$ and $V_G$. The observational data from cosmic chronometers are shown in blue dots. The obtained $\chi^2$ values for the MOG model with  $H_0=73.5$ km sec$^{-1}$ Mpc$^{-1}$  and $V_G=0$ falls beyond the  5-$\sigma$ equivalent while all other models shown in this figure are consistent with the data within 5$\sigma$.}
\label{fig:cc}  
\end{figure*}

\bgroup
\def\arraystretch{1.5}
\begin{table}
\begin{tabular}{|c|c|c|c|c|}
\hline
$V_G$ & 
\multicolumn{2}{c|}{$r_d/r_d^{\rm fid}$}&    \multicolumn{2}{c|}{$\tilde{\chi}^2$} \\
\cline{2-3} \cline{4-5}
    & $H_0=67.4$ & $H_0=73.5$ & $H_0=67.4$ & $H_0=73.5$ \\
\hline
$0$ & $0.943 \pm 0.008$ & $0.947 \pm 0.008$ & 46.72 & 84.82 \\
$0.6V_G^M$ & $0.949 \pm 0.008$ & $0.924 \pm 0.008$ &  11.72 & 37.47\\
$0.8V_G^M$ & $0.967 \pm 0.008$ & $0.925 \pm 0.008$ & 4.02 & 23.39 \\
$0.9V_G^M$ & $0.982 \pm 0.008$ & $0.928 \pm 0.008$ & 1.58 & 17.08\\
$V_G^M$ & $1.007 \pm 0.008$ & $0.934 \pm 0.008$ & 0.9 & 11.43\\
$1.1V_G^M$ & $1.047 \pm 0.008$ & $0.943 \pm 0.008$ & 4.55 & 6.70 \\
$1.2V_G^M$ & $1.138 \pm 0.010$ & $0.956 \pm 0.008$ & 28.12 &  3.39\\
\hline
\end{tabular}
\caption{Results of the statistical analysis performed with BAO data. The table shows the fixed values of $H_0$ and $V_G$ considered in each case,  and the corresponding  reduced $\chi^2$-value. The equivalent $\chi^2$ at 5$\sigma$ is  $\tilde{\chi}^2_{5\sigma}=3.73$ while for the $\Lambda$CDM fiducial models we obtain $\tilde{\chi}^2_{\Lambda{\rm CDM1}}=0.94$ and $\tilde{\chi}^2_{\Lambda{\rm CDM2}}=3.68$.}
\label{tab:chi2_BAO}
\end{table}

\begin{figure*}[t]
\centering
\includegraphics[width=0.9\textwidth, trim={0.5cm 0.5cm 0.5cm 0.5cm},clip=True]{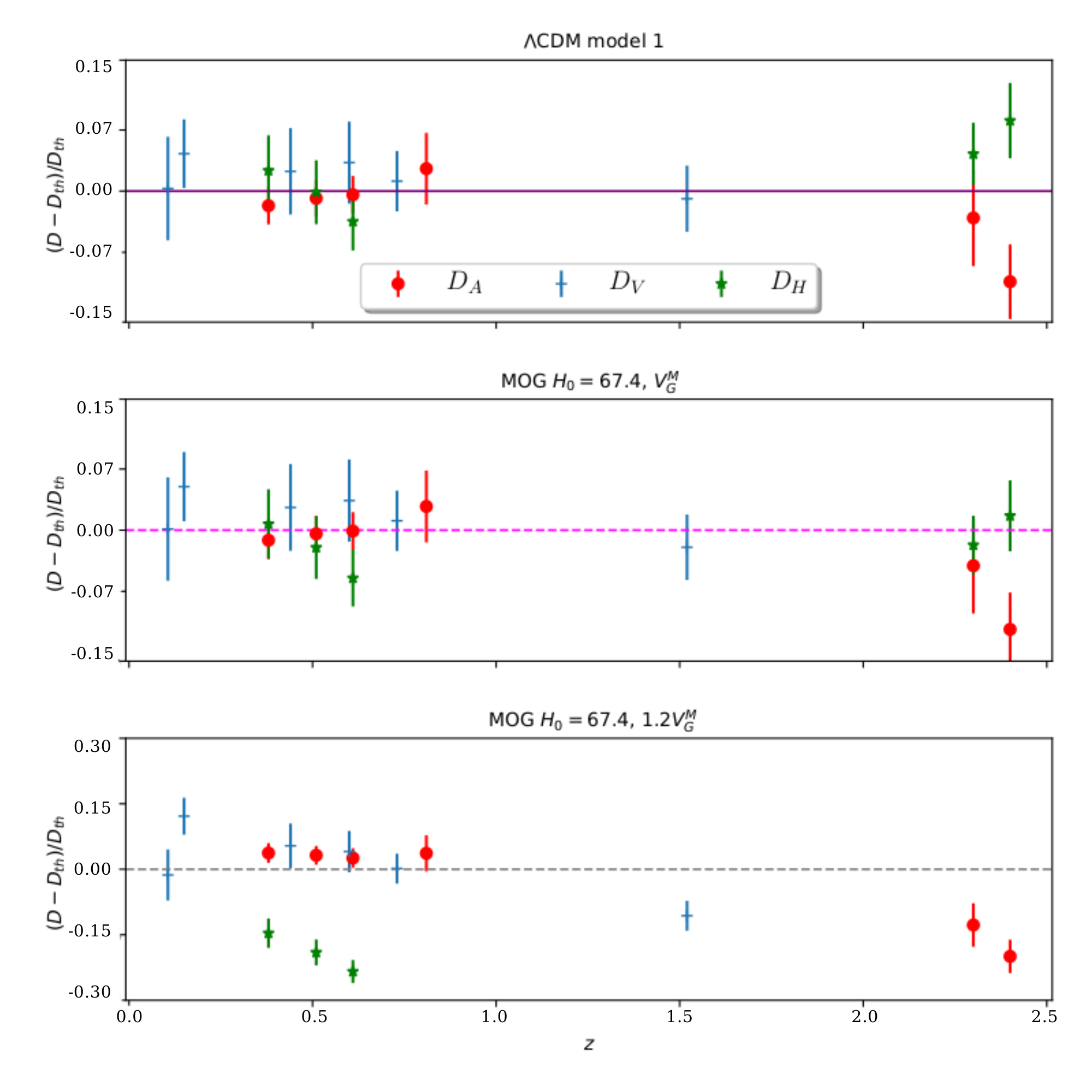}
\caption{Relative differences between the observable distance ($D_A$, $D_V$, or $D_H$) and the  corresponding theoretical 
 prediction  for the $\Lambda$CDM  model 1 and the MOG models with different values for $H_0$ and $V_G$. The fiducial  $\Lambda$CDM model 1  is shown in the upper panel and $\tilde{\chi}^2_{\Lambda{\rm CDM}}=0.94$. }
\label{fig:bao}  
\end{figure*}

\begin{figure*}[t]
\centering
\includegraphics[width=0.9\textwidth, trim={0.5cm 0.5cm 0.5cm 0.5cm},clip=True]{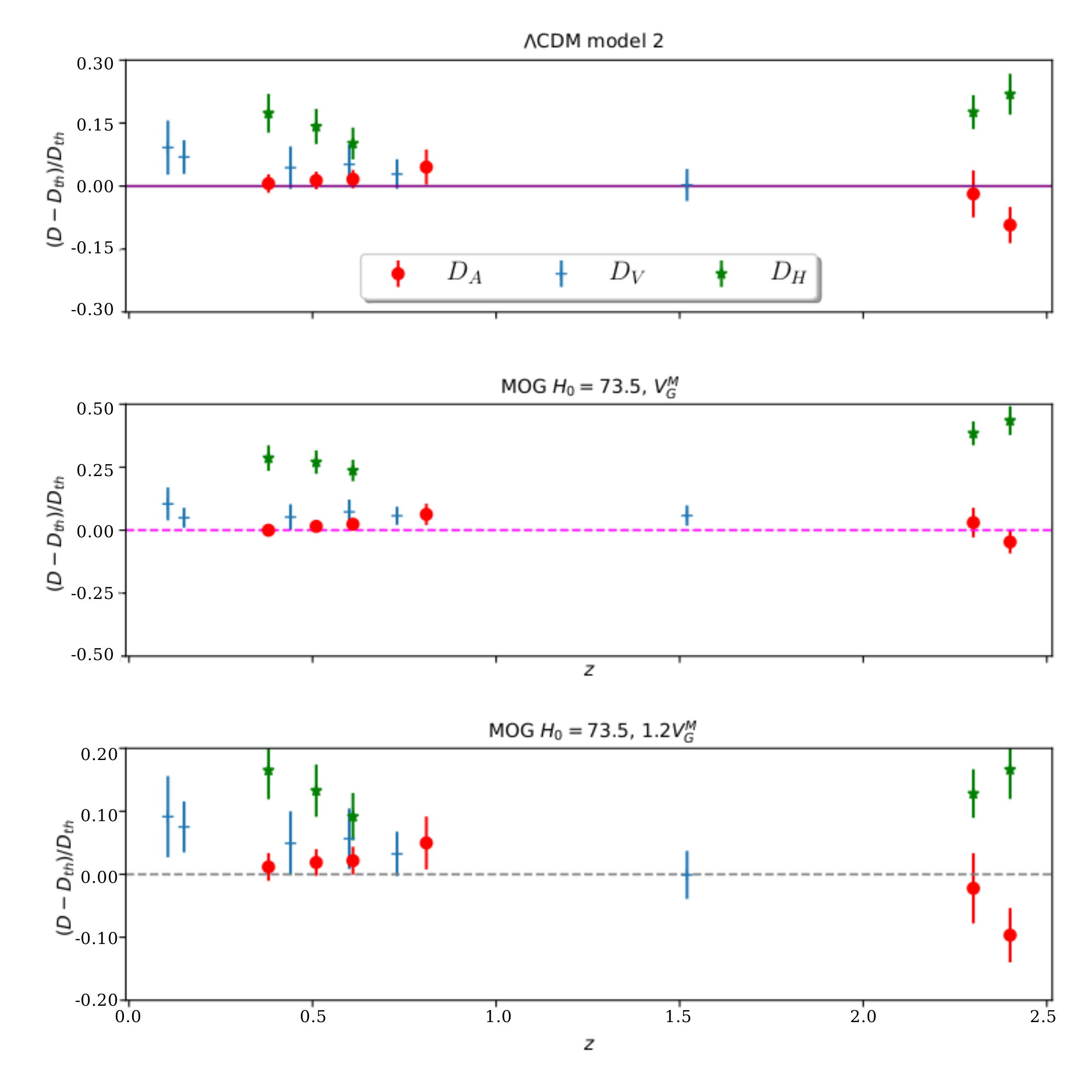}
\caption{Relative differences between the observable distance ($D_A$, $D_V$, or $D_H$) and the  corresponding theoretical 
 prediction  for the $\Lambda$CDM model 2 and the MOG models with different values for $H_0$ and $V_G$. The fiducial  $\Lambda$CDM model  is shown in the upper panel and $\tilde{\chi}^2_{\Lambda{\rm CDM}}=3.68$. }
\label{fig:bao2}  
\end{figure*}

\section{Summary and Conclusions}
\label{conclusions}
\par In this paper we have analyzed  the MOG theory in the cosmological context. We considered recent data sets from SNe Ia, BAO and CC to perform $\chi^2$ tests. We  considered models with  fixed values of $H_0$ and $V_G$, the self-interaction potential of the scalar field that represents the gravitational constant in this theory. 
\par Our results show that none of the predictions of the MOG theory considered here with $V_G=0$ is in agreement with the data, which is also consistent with previous results~\cite{MT2007,2013Galax...1...65M,2016EPJC...76..490J,2018arXiv181104445J}. Regarding the SNe Ia data, we have verified that for different non-zero values of $V_G$, 
the difference between the predictions of the MOG theory is very tiny (in fact, adding more MOG models to  Figs~\ref{fig:snia} and~\ref{fig:snia_2}  would have been confusing due to the superposition of curves). Conversely, in the case of the CC data, the predictions vary with the value of $V_G$ and $H_0$, but the large error bars  prevent it from being possible to discard more models with this data set. Furthermore, we have also shown that there is agreement between all SNe Ia and all but one CC data and the predictions of the MOG theory. 
The result is completely different in the case of the statistical analysis with the BAO data set, in which only three  models are consistent with the data within 5$\sigma$. 
 Furthermore, it should be stressed that the statistical significance  of these latter results  is greater when $H_0=64.7$  km sec$^{-1}$ Mpc$^{-1}$ is considered. 
 
 \par On the other hand, the present analysis confirms that  the nuisance parameters obtained by the Pantheon compilation assuming the $\Lambda$CDM model are also valid for the MOG theory and therefore it is likely for them to be  accurate for other non-standard cosmological models.
 
 \par  In summary, we have analysed the predictions of MOG theory for different values of the Hubble constant and of the self-interaction potential of the scalar field, and compared them to cosmological data of distinct nature, at different redshifts.  We have found that most of the studied cases are ruled out by BAO data set, although there are still some values for which the MOG theory cannot be ruled out for any of the data sets considered in this work.
 This might mean that MOG theory is still a valid alternative to the dark sector of the Universe.

\section{Acknowledgements}
The authors are supported  by the National Agency for the Promotion of Science and Technology (ANPCYT) of Argentina grant PICT-2016-0081; and grants G140 and G157 from UNLP.  S.L. acknowledges support from grant UBACYT 20020170100129BA
\par 

\bibliographystyle{apsrev}
\bibliography{testmog}

\begin{thebibliography}{84}
\expandafter\ifx\csname natexlab\endcsname\relax\def\natexlab#1{#1}\fi
\expandafter\ifx\csname bibnamefont\endcsname\relax
  \def\bibnamefont#1{#1}\fi
\expandafter\ifx\csname bibfnamefont\endcsname\relax
  \def\bibfnamefont#1{#1}\fi
\expandafter\ifx\csname citenamefont\endcsname\relax
  \def\citenamefont#1{#1}\fi
\expandafter\ifx\csname url\endcsname\relax
  \def\url#1{\texttt{#1}}\fi
\expandafter\ifx\csname urlprefix\endcsname\relax\def\urlprefix{URL }\fi
\providecommand{\bibinfo}[2]{#2}
\providecommand{\eprint}[2][]{\url{#2}}

\bibitem[{\citenamefont{{Oort}}(1932)}]{Oort1932}
\bibinfo{author}{\bibfnamefont{J.~H.} \bibnamefont{{Oort}}},
  \bibinfo{journal}{Bulletin of the Astronomical Institutes of the Netherlands}
  \textbf{\bibinfo{volume}{6}}, \bibinfo{pages}{249} (\bibinfo{year}{1932}).

\bibitem[{\citenamefont{{Zwicky}}(1937)}]{Zwicky1937}
\bibinfo{author}{\bibfnamefont{F.}~\bibnamefont{{Zwicky}}},
  \bibinfo{journal}{Astrophysical Journal} \textbf{\bibinfo{volume}{86}},
  \bibinfo{pages}{217} (\bibinfo{year}{1937}).

\bibitem[{\citenamefont{{Rubin} et~al.}(1978)\citenamefont{{Rubin}, {Ford}, and
  {Thonnard}}}]{Rubin1978}
\bibinfo{author}{\bibfnamefont{V.~C.} \bibnamefont{{Rubin}}},
  \bibinfo{author}{\bibfnamefont{J.}~\bibnamefont{{Ford}},
  \bibfnamefont{W.~K.}}, \bibnamefont{and}
  \bibinfo{author}{\bibfnamefont{N.}~\bibnamefont{{Thonnard}}},
  \bibinfo{journal}{Astrophysical Journal l} \textbf{\bibinfo{volume}{225}},
  \bibinfo{pages}{L107} (\bibinfo{year}{1978}).

\bibitem[{\citenamefont{{Rubin} et~al.}(1980)\citenamefont{{Rubin}, {Ford}, and
  {Thonnard}}}]{Rubin1980}
\bibinfo{author}{\bibfnamefont{V.~C.} \bibnamefont{{Rubin}}},
  \bibinfo{author}{\bibfnamefont{J.}~\bibnamefont{{Ford}},
  \bibfnamefont{W.~K.}}, \bibnamefont{and}
  \bibinfo{author}{\bibfnamefont{N.}~\bibnamefont{{Thonnard}}},
  \bibinfo{journal}{Astrophysical Journal} \textbf{\bibinfo{volume}{238}},
  \bibinfo{pages}{471} (\bibinfo{year}{1980}).

\bibitem[{\citenamefont{{Tyson} et~al.}(1998)\citenamefont{{Tyson},
  {Kochanski}, and {Dell'Antonio}}}]{1998ApJ...498L.107T}
\bibinfo{author}{\bibfnamefont{J.~A.} \bibnamefont{{Tyson}}},
  \bibinfo{author}{\bibfnamefont{G.~P.} \bibnamefont{{Kochanski}}},
  \bibnamefont{and} \bibinfo{author}{\bibfnamefont{I.~P.}
  \bibnamefont{{Dell'Antonio}}}, \bibinfo{journal}{Astrophys. J. Lett.}
  \textbf{\bibinfo{volume}{498}}, \bibinfo{pages}{L107} (\bibinfo{year}{1998}),
  \eprint{astro-ph/9801193}.

\bibitem[{\citenamefont{{Hasselfield} et~al.}(2013)\citenamefont{{Hasselfield},
  {Hilton}, {Marriage}, {Addison}, {Barrientos}, {Battaglia}, {Battistelli},
  {Bond}, {Crichton}, {Das} et~al.}}]{2013JCAP...07..008H}
\bibinfo{author}{\bibfnamefont{M.}~\bibnamefont{{Hasselfield}}},
  \bibinfo{author}{\bibfnamefont{M.}~\bibnamefont{{Hilton}}},
  \bibinfo{author}{\bibfnamefont{T.~A.} \bibnamefont{{Marriage}}},
  \bibinfo{author}{\bibfnamefont{G.~E.} \bibnamefont{{Addison}}},
  \bibinfo{author}{\bibfnamefont{L.~F.} \bibnamefont{{Barrientos}}},
  \bibinfo{author}{\bibfnamefont{N.}~\bibnamefont{{Battaglia}}},
  \bibinfo{author}{\bibfnamefont{E.~S.} \bibnamefont{{Battistelli}}},
  \bibinfo{author}{\bibfnamefont{J.~R.} \bibnamefont{{Bond}}},
  \bibinfo{author}{\bibfnamefont{D.}~\bibnamefont{{Crichton}}},
  \bibinfo{author}{\bibfnamefont{S.}~\bibnamefont{{Das}}},
  \bibnamefont{et~al.}, \bibinfo{journal}{Journal of Cosmology and
  Astroparticle Physics} \textbf{\bibinfo{volume}{7}}, \bibinfo{eid}{008}
  (\bibinfo{year}{2013}), \eprint{1301.0816}.

\bibitem[{\citenamefont{{Osato} et~al.}(2018)\citenamefont{{Osato}, {Flender},
  {Nagai}, {Shirasaki}, and {Yoshida}}}]{2018MNRAS.475..532O}
\bibinfo{author}{\bibfnamefont{K.}~\bibnamefont{{Osato}}},
  \bibinfo{author}{\bibfnamefont{S.}~\bibnamefont{{Flender}}},
  \bibinfo{author}{\bibfnamefont{D.}~\bibnamefont{{Nagai}}},
  \bibinfo{author}{\bibfnamefont{M.}~\bibnamefont{{Shirasaki}}},
  \bibnamefont{and}
  \bibinfo{author}{\bibfnamefont{N.}~\bibnamefont{{Yoshida}}},
  \bibinfo{journal}{Mon. Not. Roy. Astron.Soc.} \textbf{\bibinfo{volume}{475}},
  \bibinfo{pages}{532} (\bibinfo{year}{2018}), \eprint{1706.08972}.

\bibitem[{\citenamefont{{Planck Collaboration}
  et~al.}(2016)\citenamefont{{Planck Collaboration}, {Ade}, {Aghanim},
  {Arnaud}, {Ashdown}, {Aumont}, {Baccigalupi}, {Banday}, {Barreiro},
  {Bartlett} et~al.}}]{2016A&A...594A..24P}
\bibinfo{author}{\bibnamefont{{Planck Collaboration}}},
  \bibinfo{author}{\bibfnamefont{P.~A.~R.} \bibnamefont{{Ade}}},
  \bibinfo{author}{\bibfnamefont{N.}~\bibnamefont{{Aghanim}}},
  \bibinfo{author}{\bibfnamefont{M.}~\bibnamefont{{Arnaud}}},
  \bibinfo{author}{\bibfnamefont{M.}~\bibnamefont{{Ashdown}}},
  \bibinfo{author}{\bibfnamefont{J.}~\bibnamefont{{Aumont}}},
  \bibinfo{author}{\bibfnamefont{C.}~\bibnamefont{{Baccigalupi}}},
  \bibinfo{author}{\bibfnamefont{A.~J.} \bibnamefont{{Banday}}},
  \bibinfo{author}{\bibfnamefont{R.~B.} \bibnamefont{{Barreiro}}},
  \bibinfo{author}{\bibfnamefont{J.~G.} \bibnamefont{{Bartlett}}},
  \bibnamefont{et~al.}, \bibinfo{journal}{Astronomy and Astrophysics}
  \textbf{\bibinfo{volume}{594}}, \bibinfo{eid}{A24} (\bibinfo{year}{2016}),
  \eprint{1502.01597}.

\bibitem[{\citenamefont{{Sakamoto} et~al.}(2003)\citenamefont{{Sakamoto},
  {Chiba}, and {Beers}}}]{2003A&A...397..899S}
\bibinfo{author}{\bibfnamefont{T.}~\bibnamefont{{Sakamoto}}},
  \bibinfo{author}{\bibfnamefont{M.}~\bibnamefont{{Chiba}}}, \bibnamefont{and}
  \bibinfo{author}{\bibfnamefont{T.~C.} \bibnamefont{{Beers}}},
  \bibinfo{journal}{Astronomy and Astrophysics} \textbf{\bibinfo{volume}{397}},
  \bibinfo{pages}{899} (\bibinfo{year}{2003}), \eprint{astro-ph/0210508}.

\bibitem[{\citenamefont{{Xue} et~al.}(2008)\citenamefont{{Xue}, {Rix}, {Zhao},
  {Re Fiorentin}, {Naab}, {Steinmetz}, {van den Bosch}, {Beers}, {Lee}, {Bell}
  et~al.}}]{2008ApJ...684.1143X}
\bibinfo{author}{\bibfnamefont{X.~X.} \bibnamefont{{Xue}}},
  \bibinfo{author}{\bibfnamefont{H.~W.} \bibnamefont{{Rix}}},
  \bibinfo{author}{\bibfnamefont{G.}~\bibnamefont{{Zhao}}},
  \bibinfo{author}{\bibfnamefont{P.}~\bibnamefont{{Re Fiorentin}}},
  \bibinfo{author}{\bibfnamefont{T.}~\bibnamefont{{Naab}}},
  \bibinfo{author}{\bibfnamefont{M.}~\bibnamefont{{Steinmetz}}},
  \bibinfo{author}{\bibfnamefont{F.~C.} \bibnamefont{{van den Bosch}}},
  \bibinfo{author}{\bibfnamefont{T.~C.} \bibnamefont{{Beers}}},
  \bibinfo{author}{\bibfnamefont{Y.~S.} \bibnamefont{{Lee}}},
  \bibinfo{author}{\bibfnamefont{E.~F.} \bibnamefont{{Bell}}},
  \bibnamefont{et~al.}, \bibinfo{journal}{Astrophysical Journal}
  \textbf{\bibinfo{volume}{684}}, \bibinfo{pages}{1143} (\bibinfo{year}{2008}),
  \eprint{0801.1232}.

\bibitem[{\citenamefont{{Kafle} et~al.}(2014)\citenamefont{{Kafle}, {Sharma},
  {Lewis}, and {Bland-Hawthorn}}}]{2014ApJ...794...59K}
\bibinfo{author}{\bibfnamefont{P.~R.} \bibnamefont{{Kafle}}},
  \bibinfo{author}{\bibfnamefont{S.}~\bibnamefont{{Sharma}}},
  \bibinfo{author}{\bibfnamefont{G.~F.} \bibnamefont{{Lewis}}},
  \bibnamefont{and}
  \bibinfo{author}{\bibfnamefont{J.}~\bibnamefont{{Bland-Hawthorn}}},
  \bibinfo{journal}{Astrophysical Journal} \textbf{\bibinfo{volume}{794}},
  \bibinfo{eid}{59} (\bibinfo{year}{2014}), \eprint{1408.1787}.

\bibitem[{\citenamefont{{Iocco} et~al.}(2015)\citenamefont{{Iocco}, {Pato}, and
  {Bertone}}}]{2015NatPh..11..245I}
\bibinfo{author}{\bibfnamefont{F.}~\bibnamefont{{Iocco}}},
  \bibinfo{author}{\bibfnamefont{M.}~\bibnamefont{{Pato}}}, \bibnamefont{and}
  \bibinfo{author}{\bibfnamefont{G.}~\bibnamefont{{Bertone}}},
  \bibinfo{journal}{Nature Physics} \textbf{\bibinfo{volume}{11}},
  \bibinfo{pages}{245} (\bibinfo{year}{2015}), \eprint{1502.03821}.

\bibitem[{\citenamefont{{Scolnic} et~al.}(2018)\citenamefont{{Scolnic},
  {Jones}, {Rest}, {Pan}, {Chornock}, {Foley}, {Huber}, {Kessler}, {Narayan},
  {Riess} et~al.}}]{2018ApJ...859..101S}
\bibinfo{author}{\bibfnamefont{D.~M.} \bibnamefont{{Scolnic}}},
  \bibinfo{author}{\bibfnamefont{D.~O.} \bibnamefont{{Jones}}},
  \bibinfo{author}{\bibfnamefont{A.}~\bibnamefont{{Rest}}},
  \bibinfo{author}{\bibfnamefont{Y.~C.} \bibnamefont{{Pan}}},
  \bibinfo{author}{\bibfnamefont{R.}~\bibnamefont{{Chornock}}},
  \bibinfo{author}{\bibfnamefont{R.~J.} \bibnamefont{{Foley}}},
  \bibinfo{author}{\bibfnamefont{M.~E.} \bibnamefont{{Huber}}},
  \bibinfo{author}{\bibfnamefont{R.}~\bibnamefont{{Kessler}}},
  \bibinfo{author}{\bibfnamefont{G.}~\bibnamefont{{Narayan}}},
  \bibinfo{author}{\bibfnamefont{A.~G.} \bibnamefont{{Riess}}},
  \bibnamefont{et~al.}, \bibinfo{journal}{Astrophysical Journal}
  \textbf{\bibinfo{volume}{859}}, \bibinfo{eid}{101} (\bibinfo{year}{2018}),
  \eprint{1710.00845}.

\bibitem[{\citenamefont{{Planck Collaboration}
  et~al.}(2018)\citenamefont{{Planck Collaboration}, {Aghanim}, {Akrami},
  {Ashdown}, {Aumont}, {Baccigalupi}, {Ballardini}, {Banday}, {Barreiro},
  {Bartolo} et~al.}}]{Planckcosmo2018}
\bibinfo{author}{\bibnamefont{{Planck Collaboration}}},
  \bibinfo{author}{\bibfnamefont{N.}~\bibnamefont{{Aghanim}}},
  \bibinfo{author}{\bibfnamefont{Y.}~\bibnamefont{{Akrami}}},
  \bibinfo{author}{\bibfnamefont{M.}~\bibnamefont{{Ashdown}}},
  \bibinfo{author}{\bibfnamefont{J.}~\bibnamefont{{Aumont}}},
  \bibinfo{author}{\bibfnamefont{C.}~\bibnamefont{{Baccigalupi}}},
  \bibinfo{author}{\bibfnamefont{M.}~\bibnamefont{{Ballardini}}},
  \bibinfo{author}{\bibfnamefont{A.~J.} \bibnamefont{{Banday}}},
  \bibinfo{author}{\bibfnamefont{R.~B.} \bibnamefont{{Barreiro}}},
  \bibinfo{author}{\bibfnamefont{N.}~\bibnamefont{{Bartolo}}},
  \bibnamefont{et~al.}, \bibinfo{journal}{arXiv e-prints}
  \bibinfo{eid}{arXiv:1807.06209} (\bibinfo{year}{2018}), \eprint{1807.06209}.

\bibitem[{\citenamefont{Ross et~al.}(2015)\citenamefont{Ross, Samushia,
  Howlett, Percival, Burden, and Manera}}]{SDSS_DR7}
\bibinfo{author}{\bibfnamefont{A.~J.} \bibnamefont{Ross}},
  \bibinfo{author}{\bibfnamefont{L.}~\bibnamefont{Samushia}},
  \bibinfo{author}{\bibfnamefont{C.}~\bibnamefont{Howlett}},
  \bibinfo{author}{\bibfnamefont{W.~J.} \bibnamefont{Percival}},
  \bibinfo{author}{\bibfnamefont{A.}~\bibnamefont{Burden}}, \bibnamefont{and}
  \bibinfo{author}{\bibfnamefont{M.}~\bibnamefont{Manera}},
  \bibinfo{journal}{Monthly Notices of the Royal Astronomical Society}
  \textbf{\bibinfo{volume}{449}}, \bibinfo{pages}{835–847}
  (\bibinfo{year}{2015}), ISSN \bibinfo{issn}{0035-8711},
  \urlprefix\url{http://dx.doi.org/10.1093/mnras/stv154}.

\bibitem[{\citenamefont{Alam et~al.}(2017)\citenamefont{Alam, Ata, Bailey,
  Beutler, Bizyaev, Blazek, Bolton, Brownstein, Burden, Chuang et~al.}}]{BOSS}
\bibinfo{author}{\bibfnamefont{S.}~\bibnamefont{Alam}},
  \bibinfo{author}{\bibfnamefont{M.}~\bibnamefont{Ata}},
  \bibinfo{author}{\bibfnamefont{S.}~\bibnamefont{Bailey}},
  \bibinfo{author}{\bibfnamefont{F.}~\bibnamefont{Beutler}},
  \bibinfo{author}{\bibfnamefont{D.}~\bibnamefont{Bizyaev}},
  \bibinfo{author}{\bibfnamefont{J.~A.} \bibnamefont{Blazek}},
  \bibinfo{author}{\bibfnamefont{A.~S.} \bibnamefont{Bolton}},
  \bibinfo{author}{\bibfnamefont{J.~R.} \bibnamefont{Brownstein}},
  \bibinfo{author}{\bibfnamefont{A.}~\bibnamefont{Burden}},
  \bibinfo{author}{\bibfnamefont{C.-H.} \bibnamefont{Chuang}},
  \bibnamefont{et~al.}, \bibinfo{journal}{Monthly Notices of the Royal
  Astronomical Society} \textbf{\bibinfo{volume}{470}},
  \bibinfo{pages}{2617–2652} (\bibinfo{year}{2017}), ISSN
  \bibinfo{issn}{1365-2966},
  \urlprefix\url{http://dx.doi.org/10.1093/mnras/stx721}.

\bibitem[{\citenamefont{Abbott et~al.}(2018)\citenamefont{Abbott, Abdalla,
  Alarcon, Allam, Andrade-Oliveira, Annis, Avila, Banerji, Banik, Bechtol
  et~al.}}]{DES_Y1}
\bibinfo{author}{\bibfnamefont{T.~M.~C.} \bibnamefont{Abbott}},
  \bibinfo{author}{\bibfnamefont{F.~B.} \bibnamefont{Abdalla}},
  \bibinfo{author}{\bibfnamefont{A.}~\bibnamefont{Alarcon}},
  \bibinfo{author}{\bibfnamefont{S.}~\bibnamefont{Allam}},
  \bibinfo{author}{\bibfnamefont{F.}~\bibnamefont{Andrade-Oliveira}},
  \bibinfo{author}{\bibfnamefont{J.}~\bibnamefont{Annis}},
  \bibinfo{author}{\bibfnamefont{S.}~\bibnamefont{Avila}},
  \bibinfo{author}{\bibfnamefont{M.}~\bibnamefont{Banerji}},
  \bibinfo{author}{\bibfnamefont{N.}~\bibnamefont{Banik}},
  \bibinfo{author}{\bibfnamefont{K.}~\bibnamefont{Bechtol}},
  \bibnamefont{et~al.}, \bibinfo{journal}{Monthly Notices of the Royal
  Astronomical Society} \textbf{\bibinfo{volume}{483}},
  \bibinfo{pages}{4866–4883} (\bibinfo{year}{2018}), ISSN
  \bibinfo{issn}{1365-2966},
  \urlprefix\url{http://dx.doi.org/10.1093/mnras/sty3351}.

\bibitem[{\citenamefont{{Nuza} et~al.}(2013)\citenamefont{{Nuza},
  {S{\'a}nchez}, {Prada}, {Klypin}, {Schlegel}, {Gottl{\"o}ber},
  {Montero-Dorta}, {Manera}, {McBride}, {Ross} et~al.}}]{2013MNRAS.432..743N}
\bibinfo{author}{\bibfnamefont{S.~E.} \bibnamefont{{Nuza}}},
  \bibinfo{author}{\bibfnamefont{A.~G.} \bibnamefont{{S{\'a}nchez}}},
  \bibinfo{author}{\bibfnamefont{F.}~\bibnamefont{{Prada}}},
  \bibinfo{author}{\bibfnamefont{A.}~\bibnamefont{{Klypin}}},
  \bibinfo{author}{\bibfnamefont{D.~J.} \bibnamefont{{Schlegel}}},
  \bibinfo{author}{\bibfnamefont{S.}~\bibnamefont{{Gottl{\"o}ber}}},
  \bibinfo{author}{\bibfnamefont{A.~D.} \bibnamefont{{Montero-Dorta}}},
  \bibinfo{author}{\bibfnamefont{M.}~\bibnamefont{{Manera}}},
  \bibinfo{author}{\bibfnamefont{C.~K.} \bibnamefont{{McBride}}},
  \bibinfo{author}{\bibfnamefont{A.~J.} \bibnamefont{{Ross}}},
  \bibnamefont{et~al.}, \bibinfo{journal}{Mon. Not. Roy. Astron.Soc.}
  \textbf{\bibinfo{volume}{432}}, \bibinfo{pages}{743} (\bibinfo{year}{2013}),
  \eprint{1202.6057}.

\bibitem[{\citenamefont{{Schumann}}(2019)}]{2019JPhG...46j3003S}
\bibinfo{author}{\bibfnamefont{M.}~\bibnamefont{{Schumann}}},
  \bibinfo{journal}{Journal of Physics G Nuclear Physics}
  \textbf{\bibinfo{volume}{46}}, \bibinfo{pages}{103003}
  (\bibinfo{year}{2019}), \eprint{1903.03026}.

\bibitem[{\citenamefont{{Milgrom}}(1983)}]{1983ApJ...270..365M}
\bibinfo{author}{\bibfnamefont{M.}~\bibnamefont{{Milgrom}}},
  \bibinfo{journal}{Astrophysical Journal} \textbf{\bibinfo{volume}{270}},
  \bibinfo{pages}{365} (\bibinfo{year}{1983}).

\bibitem[{\citenamefont{{Bekenstein}}(2004)}]{2004PhRvD..70h3509B}
\bibinfo{author}{\bibfnamefont{J.~D.} \bibnamefont{{Bekenstein}}},
  \bibinfo{journal}{\prd} \textbf{\bibinfo{volume}{70}}, \bibinfo{eid}{083509}
  (\bibinfo{year}{2004}), \eprint{astro-ph/0403694}.

\bibitem[{\citenamefont{{Clowe}
  et~al.}(2006{\natexlab{a}})\citenamefont{{Clowe}, {Brada{\v{c}}}, {Gonzalez},
  {Markevitch}, {Randall}, {Jones}, and {Zaritsky}}}]{Clowe2006}
\bibinfo{author}{\bibfnamefont{D.}~\bibnamefont{{Clowe}}},
  \bibinfo{author}{\bibfnamefont{M.}~\bibnamefont{{Brada{\v{c}}}}},
  \bibinfo{author}{\bibfnamefont{A.~H.} \bibnamefont{{Gonzalez}}},
  \bibinfo{author}{\bibfnamefont{M.}~\bibnamefont{{Markevitch}}},
  \bibinfo{author}{\bibfnamefont{S.~W.} \bibnamefont{{Randall}}},
  \bibinfo{author}{\bibfnamefont{C.}~\bibnamefont{{Jones}}}, \bibnamefont{and}
  \bibinfo{author}{\bibfnamefont{D.}~\bibnamefont{{Zaritsky}}},
  \bibinfo{journal}{Astrophysical Journal l} \textbf{\bibinfo{volume}{648}},
  \bibinfo{pages}{L109} (\bibinfo{year}{2006}{\natexlab{a}}),
  \eprint{astro-ph/0608407}.

\bibitem[{\citenamefont{{Moffat}}(2006)}]{2006JCAP...03..004M}
\bibinfo{author}{\bibfnamefont{J.~W.} \bibnamefont{{Moffat}}},
  \bibinfo{journal}{Journal of Cosmology and Gravitation}
  \textbf{\bibinfo{volume}{3}}, \bibinfo{eid}{004} (\bibinfo{year}{2006}),
  \eprint{gr-qc/0506021}.

\bibitem[{\citenamefont{{Moffat} and {Toth}}(2008)}]{2008ApJ...680.1158M}
\bibinfo{author}{\bibfnamefont{J.~W.} \bibnamefont{{Moffat}}} \bibnamefont{and}
  \bibinfo{author}{\bibfnamefont{V.~T.} \bibnamefont{{Toth}}},
  \bibinfo{journal}{Astrophysical Journal} \textbf{\bibinfo{volume}{680}},
  \bibinfo{eid}{1158-1161} (\bibinfo{year}{2008}), \eprint{0708.1935}.

\bibitem[{\citenamefont{{Moffat} and {Rahvar}}(2014)}]{2014MNRAS.441.3724M}
\bibinfo{author}{\bibfnamefont{J.~W.} \bibnamefont{{Moffat}}} \bibnamefont{and}
  \bibinfo{author}{\bibfnamefont{S.}~\bibnamefont{{Rahvar}}},
  \bibinfo{journal}{MNRAS} \textbf{\bibinfo{volume}{441}},
  \bibinfo{pages}{3724} (\bibinfo{year}{2014}), \eprint{1309.5077}.

\bibitem[{\citenamefont{{Moffat} and {Zhoolideh
  Haghighi}}(2017)}]{2017EPJP..132..417M}
\bibinfo{author}{\bibfnamefont{J.~W.} \bibnamefont{{Moffat}}} \bibnamefont{and}
  \bibinfo{author}{\bibfnamefont{M.~H.} \bibnamefont{{Zhoolideh Haghighi}}},
  \bibinfo{journal}{European Physical Journal Plus}
  \textbf{\bibinfo{volume}{132}}, \bibinfo{pages}{417} (\bibinfo{year}{2017}).

\bibitem[{\citenamefont{{Moffat} and {Rahvar}}(2013)}]{2013MNRAS.436.1439M}
\bibinfo{author}{\bibfnamefont{J.~W.} \bibnamefont{{Moffat}}} \bibnamefont{and}
  \bibinfo{author}{\bibfnamefont{S.}~\bibnamefont{{Rahvar}}},
  \bibinfo{journal}{MNRAS} \textbf{\bibinfo{volume}{436}},
  \bibinfo{pages}{1439} (\bibinfo{year}{2013}), \eprint{1306.6383}.

\bibitem[{\citenamefont{{Zhoolideh Haghighi} and
  {Rahvar}}(2017)}]{2017MNRAS.468.4048Z}
\bibinfo{author}{\bibfnamefont{M.~H.} \bibnamefont{{Zhoolideh Haghighi}}}
  \bibnamefont{and} \bibinfo{author}{\bibfnamefont{S.}~\bibnamefont{{Rahvar}}},
  \bibinfo{journal}{MNRAS} \textbf{\bibinfo{volume}{468}},
  \bibinfo{pages}{4048} (\bibinfo{year}{2017}), \eprint{1609.07851}.

\bibitem[{\citenamefont{{Negrelli} et~al.}(2018)\citenamefont{{Negrelli},
  {Benito}, {Landau}, {Iocco}, and {Kraiselburd}}}]{Negrelli2018}
\bibinfo{author}{\bibfnamefont{C.}~\bibnamefont{{Negrelli}}},
  \bibinfo{author}{\bibfnamefont{M.}~\bibnamefont{{Benito}}},
  \bibinfo{author}{\bibfnamefont{S.}~\bibnamefont{{Landau}}},
  \bibinfo{author}{\bibfnamefont{F.}~\bibnamefont{{Iocco}}}, \bibnamefont{and}
  \bibinfo{author}{\bibfnamefont{L.}~\bibnamefont{{Kraiselburd}}},
  \bibinfo{journal}{\prd} \textbf{\bibinfo{volume}{98}}, \bibinfo{eid}{104061}
  (\bibinfo{year}{2018}), \eprint{1810.07200}.

\bibitem[{\citenamefont{{Clowe}
  et~al.}(2006{\natexlab{b}})\citenamefont{{Clowe}, {Brada{\v c}}, {Gonzalez},
  {Markevitch}, {Randall}, {Jones}, and {Zaritsky}}}]{2006ApJ...648L.109C}
\bibinfo{author}{\bibfnamefont{D.}~\bibnamefont{{Clowe}}},
  \bibinfo{author}{\bibfnamefont{M.}~\bibnamefont{{Brada{\v c}}}},
  \bibinfo{author}{\bibfnamefont{A.~H.} \bibnamefont{{Gonzalez}}},
  \bibinfo{author}{\bibfnamefont{M.}~\bibnamefont{{Markevitch}}},
  \bibinfo{author}{\bibfnamefont{S.~W.} \bibnamefont{{Randall}}},
  \bibinfo{author}{\bibfnamefont{C.}~\bibnamefont{{Jones}}}, \bibnamefont{and}
  \bibinfo{author}{\bibfnamefont{D.}~\bibnamefont{{Zaritsky}}},
  \bibinfo{journal}{Astrophysical Journal Letters}
  \textbf{\bibinfo{volume}{648}}, \bibinfo{pages}{L109}
  (\bibinfo{year}{2006}{\natexlab{b}}), \eprint{astro-ph/0608407}.

\bibitem[{\citenamefont{{Brownstein} and {Moffat}}(2007)}]{2007MNRAS.382...29B}
\bibinfo{author}{\bibfnamefont{J.~R.} \bibnamefont{{Brownstein}}}
  \bibnamefont{and} \bibinfo{author}{\bibfnamefont{J.~W.}
  \bibnamefont{{Moffat}}}, \bibinfo{journal}{MNRAS}
  \textbf{\bibinfo{volume}{382}}, \bibinfo{pages}{29} (\bibinfo{year}{2007}),
  \eprint{astro-ph/0702146}.

\bibitem[{\citenamefont{{Israel} and {Moffat}}(2016)}]{2016arXiv160609128I}
\bibinfo{author}{\bibfnamefont{N.~S.} \bibnamefont{{Israel}}} \bibnamefont{and}
  \bibinfo{author}{\bibfnamefont{J.~W.} \bibnamefont{{Moffat}}},
  \bibinfo{journal}{ArXiv e-prints}  (\bibinfo{year}{2016}),
  \eprint{1606.09128}.

\bibitem[{\citenamefont{{Nieuwenhuizen}
  et~al.}(2018)\citenamefont{{Nieuwenhuizen}, {Morandi}, and
  {Limousin}}}]{2018MNRAS.tmp..370N}
\bibinfo{author}{\bibfnamefont{T.~M.} \bibnamefont{{Nieuwenhuizen}}},
  \bibinfo{author}{\bibfnamefont{A.}~\bibnamefont{{Morandi}}},
  \bibnamefont{and}
  \bibinfo{author}{\bibfnamefont{M.}~\bibnamefont{{Limousin}}},
  \bibinfo{journal}{Mon. Not. Roy. Astron.Soc.}  (\bibinfo{year}{2018}),
  \eprint{1802.04891}.

\bibitem[{\citenamefont{{Boran} et~al.}(2018)\citenamefont{{Boran}, {Desai},
  {Kahya}, and {Woodard}}}]{Boran2018}
\bibinfo{author}{\bibfnamefont{S.}~\bibnamefont{{Boran}}},
  \bibinfo{author}{\bibfnamefont{S.}~\bibnamefont{{Desai}}},
  \bibinfo{author}{\bibfnamefont{E.~O.} \bibnamefont{{Kahya}}},
  \bibnamefont{and} \bibinfo{author}{\bibfnamefont{R.~P.}
  \bibnamefont{{Woodard}}}, \bibinfo{journal}{\prd}
  \textbf{\bibinfo{volume}{97}}, \bibinfo{eid}{041501} (\bibinfo{year}{2018}),
  \eprint{1710.06168}.

\bibitem[{\citenamefont{{Green} et~al.}(2018)\citenamefont{{Green}, {Moffat},
  and {Toth}}}]{Green2018}
\bibinfo{author}{\bibfnamefont{M.~A.} \bibnamefont{{Green}}},
  \bibinfo{author}{\bibfnamefont{J.~W.} \bibnamefont{{Moffat}}},
  \bibnamefont{and} \bibinfo{author}{\bibfnamefont{V.~T.}
  \bibnamefont{{Toth}}}, \bibinfo{journal}{Physics Letters B}
  \textbf{\bibinfo{volume}{780}}, \bibinfo{pages}{300} (\bibinfo{year}{2018}),
  \eprint{1710.11177}.

\bibitem[{\citenamefont{{Schmidt} et~al.}(1998)\citenamefont{{Schmidt},
  {Suntzeff}, {Phillips}, {Schommer}, {Clocchiatti}, {Kirshner}, {Garnavich},
  {Challis}, {Leibundgut}, {Spyromilio} et~al.}}]{Schmidt1998}
\bibinfo{author}{\bibfnamefont{B.~P.} \bibnamefont{{Schmidt}}},
  \bibinfo{author}{\bibfnamefont{N.~B.} \bibnamefont{{Suntzeff}}},
  \bibinfo{author}{\bibfnamefont{M.~M.} \bibnamefont{{Phillips}}},
  \bibinfo{author}{\bibfnamefont{R.~A.} \bibnamefont{{Schommer}}},
  \bibinfo{author}{\bibfnamefont{A.}~\bibnamefont{{Clocchiatti}}},
  \bibinfo{author}{\bibfnamefont{R.~P.} \bibnamefont{{Kirshner}}},
  \bibinfo{author}{\bibfnamefont{P.}~\bibnamefont{{Garnavich}}},
  \bibinfo{author}{\bibfnamefont{P.}~\bibnamefont{{Challis}}},
  \bibinfo{author}{\bibfnamefont{B.}~\bibnamefont{{Leibundgut}}},
  \bibinfo{author}{\bibfnamefont{J.}~\bibnamefont{{Spyromilio}}},
  \bibnamefont{et~al.}, \bibinfo{journal}{Astrophysical Journal}
  \textbf{\bibinfo{volume}{507}}, \bibinfo{pages}{46} (\bibinfo{year}{1998}),
  \eprint{astro-ph/9805200}.

\bibitem[{\citenamefont{{Tsujikawa}}(2011)}]{Tsujikawa2011}
\bibinfo{author}{\bibfnamefont{S.}~\bibnamefont{{Tsujikawa}}},
  \emph{\bibinfo{title}{{Dark Energy: Investigation and Modeling}}}, vol.
  \bibinfo{volume}{370} of \emph{\bibinfo{series}{Astrophysics and Space
  Science Library}} (\bibinfo{year}{2011}).

\bibitem[{\citenamefont{{De Felice} and {Tsujikawa}}(2010)}]{DeFelice2010}
\bibinfo{author}{\bibfnamefont{A.}~\bibnamefont{{De Felice}}} \bibnamefont{and}
  \bibinfo{author}{\bibfnamefont{S.}~\bibnamefont{{Tsujikawa}}},
  \bibinfo{journal}{Living Reviews in Relativity}
  \textbf{\bibinfo{volume}{13}}, \bibinfo{eid}{3} (\bibinfo{year}{2010}),
  \eprint{1002.4928}.

\bibitem[{\citenamefont{{Moffat} and {Toth}}(2009)}]{2009CQGra..26h5002M}
\bibinfo{author}{\bibfnamefont{J.~W.} \bibnamefont{{Moffat}}} \bibnamefont{and}
  \bibinfo{author}{\bibfnamefont{V.~T.} \bibnamefont{{Toth}}},
  \bibinfo{journal}{Classical and Quantum Gravity}
  \textbf{\bibinfo{volume}{26}}, \bibinfo{eid}{085002} (\bibinfo{year}{2009}),
  \eprint{0712.1796}.

\bibitem[{\citenamefont{{Moffat} and {Toth}}(2007)}]{MT2007}
\bibinfo{author}{\bibfnamefont{J.~W.} \bibnamefont{{Moffat}}} \bibnamefont{and}
  \bibinfo{author}{\bibfnamefont{V.~T.} \bibnamefont{{Toth}}},
  \bibinfo{journal}{arXiv e-prints}  (\bibinfo{year}{2007}),
  \eprint{0710.0364}.

\bibitem[{\citenamefont{{Moffat} and {Toth}}(2013)}]{2013Galax...1...65M}
\bibinfo{author}{\bibfnamefont{J.}~\bibnamefont{{Moffat}}} \bibnamefont{and}
  \bibinfo{author}{\bibfnamefont{V.}~\bibnamefont{{Toth}}},
  \bibinfo{journal}{Galaxies} \textbf{\bibinfo{volume}{1}}, \bibinfo{pages}{65}
  (\bibinfo{year}{2013}).

\bibitem[{\citenamefont{{Toth}}(2010)}]{2010arXiv1011.5174T}
\bibinfo{author}{\bibfnamefont{V.~T.} \bibnamefont{{Toth}}},
  \bibinfo{journal}{arXiv e-prints} \bibinfo{eid}{arXiv:1011.5174}
  (\bibinfo{year}{2010}), \eprint{1011.5174}.

\bibitem[{\citenamefont{{Jamali} and {Roshan}}(2016)}]{2016EPJC...76..490J}
\bibinfo{author}{\bibfnamefont{S.}~\bibnamefont{{Jamali}}} \bibnamefont{and}
  \bibinfo{author}{\bibfnamefont{M.}~\bibnamefont{{Roshan}}},
  \bibinfo{journal}{European Physical Journal C} \textbf{\bibinfo{volume}{76}},
  \bibinfo{eid}{490} (\bibinfo{year}{2016}), \eprint{1608.06251}.

\bibitem[{\citenamefont{{Jamali} et~al.}(2018)\citenamefont{{Jamali}, {Roshan},
  and {Amendola}}}]{2018arXiv181104445J}
\bibinfo{author}{\bibfnamefont{S.}~\bibnamefont{{Jamali}}},
  \bibinfo{author}{\bibfnamefont{M.}~\bibnamefont{{Roshan}}}, \bibnamefont{and}
  \bibinfo{author}{\bibfnamefont{L.}~\bibnamefont{{Amendola}}},
  \bibinfo{journal}{arXiv e-prints} \bibinfo{eid}{arXiv:1811.04445}
  (\bibinfo{year}{2018}), \eprint{1811.04445}.

\bibitem[{\citenamefont{{Jimenez} and {Loeb}}(2002)}]{2002ApJ...573...37J}
\bibinfo{author}{\bibfnamefont{R.}~\bibnamefont{{Jimenez}}} \bibnamefont{and}
  \bibinfo{author}{\bibfnamefont{A.}~\bibnamefont{{Loeb}}},
  \bibinfo{journal}{Astrophysical Journal} \textbf{\bibinfo{volume}{573}},
  \bibinfo{pages}{37} (\bibinfo{year}{2002}), \eprint{astro-ph/0106145}.

\bibitem[{\citenamefont{{Simon} et~al.}(2005)\citenamefont{{Simon}, {Verde},
  and {Jimenez}}}]{simon05}
\bibinfo{author}{\bibfnamefont{J.}~\bibnamefont{{Simon}}},
  \bibinfo{author}{\bibfnamefont{L.}~\bibnamefont{{Verde}}}, \bibnamefont{and}
  \bibinfo{author}{\bibfnamefont{R.}~\bibnamefont{{Jimenez}}},
  \bibinfo{journal}{\prd} \textbf{\bibinfo{volume}{71}}, \bibinfo{eid}{123001}
  (\bibinfo{year}{2005}), \eprint{astro-ph/0412269}.

\bibitem[{\citenamefont{{Abraham} et~al.}(2004)\citenamefont{{Abraham},
  {Glazebrook}, {McCarthy}, {Crampton}, {Murowinski}, {J{\o}rgensen}, {Roth},
  {Hook}, {Savaglio}, {Chen} et~al.}}]{Abraham2004}
\bibinfo{author}{\bibfnamefont{R.~G.} \bibnamefont{{Abraham}}},
  \bibinfo{author}{\bibfnamefont{K.}~\bibnamefont{{Glazebrook}}},
  \bibinfo{author}{\bibfnamefont{P.~J.} \bibnamefont{{McCarthy}}},
  \bibinfo{author}{\bibfnamefont{D.}~\bibnamefont{{Crampton}}},
  \bibinfo{author}{\bibfnamefont{R.}~\bibnamefont{{Murowinski}}},
  \bibinfo{author}{\bibfnamefont{I.}~\bibnamefont{{J{\o}rgensen}}},
  \bibinfo{author}{\bibfnamefont{K.}~\bibnamefont{{Roth}}},
  \bibinfo{author}{\bibfnamefont{I.~M.} \bibnamefont{{Hook}}},
  \bibinfo{author}{\bibfnamefont{S.}~\bibnamefont{{Savaglio}}},
  \bibinfo{author}{\bibfnamefont{H.-W.} \bibnamefont{{Chen}}},
  \bibnamefont{et~al.}, \bibinfo{journal}{Astronomical Journal}
  \textbf{\bibinfo{volume}{127}}, \bibinfo{pages}{2455} (\bibinfo{year}{2004}),
  \eprint{astro-ph/0402436}.

\bibitem[{\citenamefont{{Dunlop} et~al.}(1996)\citenamefont{{Dunlop},
  {Peacock}, {Spinrad}, {Dey}, {Jimenez}, {Stern}, and
  {Windhorst}}}]{Dunlop1996}
\bibinfo{author}{\bibfnamefont{J.}~\bibnamefont{{Dunlop}}},
  \bibinfo{author}{\bibfnamefont{J.}~\bibnamefont{{Peacock}}},
  \bibinfo{author}{\bibfnamefont{H.}~\bibnamefont{{Spinrad}}},
  \bibinfo{author}{\bibfnamefont{A.}~\bibnamefont{{Dey}}},
  \bibinfo{author}{\bibfnamefont{R.}~\bibnamefont{{Jimenez}}},
  \bibinfo{author}{\bibfnamefont{D.}~\bibnamefont{{Stern}}}, \bibnamefont{and}
  \bibinfo{author}{\bibfnamefont{R.}~\bibnamefont{{Windhorst}}},
  \bibinfo{journal}{\nat} \textbf{\bibinfo{volume}{381}}, \bibinfo{pages}{581}
  (\bibinfo{year}{1996}).

\bibitem[{\citenamefont{{Nolan} et~al.}(2003)\citenamefont{{Nolan}, {Dunlop},
  {Jimenez}, and {Heavens}}}]{Nolan2003}
\bibinfo{author}{\bibfnamefont{L.~A.} \bibnamefont{{Nolan}}},
  \bibinfo{author}{\bibfnamefont{J.~S.} \bibnamefont{{Dunlop}}},
  \bibinfo{author}{\bibfnamefont{R.}~\bibnamefont{{Jimenez}}},
  \bibnamefont{and} \bibinfo{author}{\bibfnamefont{A.~F.}
  \bibnamefont{{Heavens}}}, \bibinfo{journal}{\mnras}
  \textbf{\bibinfo{volume}{341}}, \bibinfo{pages}{464} (\bibinfo{year}{2003}),
  \eprint{astro-ph/0103450}.

\bibitem[{\citenamefont{{Spinrad} et~al.}(1997)\citenamefont{{Spinrad}, {Dey},
  {Stern}, {Dunlop}, {Peacock}, {Jimenez}, and {Windhorst}}}]{Spinrad1997}
\bibinfo{author}{\bibfnamefont{H.}~\bibnamefont{{Spinrad}}},
  \bibinfo{author}{\bibfnamefont{A.}~\bibnamefont{{Dey}}},
  \bibinfo{author}{\bibfnamefont{D.}~\bibnamefont{{Stern}}},
  \bibinfo{author}{\bibfnamefont{J.}~\bibnamefont{{Dunlop}}},
  \bibinfo{author}{\bibfnamefont{J.}~\bibnamefont{{Peacock}}},
  \bibinfo{author}{\bibfnamefont{R.}~\bibnamefont{{Jimenez}}},
  \bibnamefont{and}
  \bibinfo{author}{\bibfnamefont{R.}~\bibnamefont{{Windhorst}}},
  \bibinfo{journal}{Astrophysical Journal} \textbf{\bibinfo{volume}{484}},
  \bibinfo{pages}{581} (\bibinfo{year}{1997}), \eprint{astro-ph/9702233}.

\bibitem[{\citenamefont{{Treu} et~al.}(1999)\citenamefont{{Treu}, {Stiavelli},
  {Casertano}, {M{\o}ller}, and {Bertin}}}]{Treu1999}
\bibinfo{author}{\bibfnamefont{T.}~\bibnamefont{{Treu}}},
  \bibinfo{author}{\bibfnamefont{M.}~\bibnamefont{{Stiavelli}}},
  \bibinfo{author}{\bibfnamefont{S.}~\bibnamefont{{Casertano}}},
  \bibinfo{author}{\bibfnamefont{P.}~\bibnamefont{{M{\o}ller}}},
  \bibnamefont{and} \bibinfo{author}{\bibfnamefont{G.}~\bibnamefont{{Bertin}}},
  \bibinfo{journal}{\mnras} \textbf{\bibinfo{volume}{308}},
  \bibinfo{pages}{1037} (\bibinfo{year}{1999}), \eprint{astro-ph/9904327}.

\bibitem[{\citenamefont{{Treu} et~al.}(2001)\citenamefont{{Treu}, {Stiavelli},
  {M{\o}ller}, {Casertano}, and {Bertin}}}]{Treu2001}
\bibinfo{author}{\bibfnamefont{T.}~\bibnamefont{{Treu}}},
  \bibinfo{author}{\bibfnamefont{M.}~\bibnamefont{{Stiavelli}}},
  \bibinfo{author}{\bibfnamefont{P.}~\bibnamefont{{M{\o}ller}}},
  \bibinfo{author}{\bibfnamefont{S.}~\bibnamefont{{Casertano}}},
  \bibnamefont{and} \bibinfo{author}{\bibfnamefont{G.}~\bibnamefont{{Bertin}}},
  \bibinfo{journal}{\mnras} \textbf{\bibinfo{volume}{326}},
  \bibinfo{pages}{221} (\bibinfo{year}{2001}), \eprint{astro-ph/0104177}.

\bibitem[{\citenamefont{{Treu} et~al.}(2002)\citenamefont{{Treu}, {Stiavelli},
  {Casertano}, {M{\o}ller}, and {Bertin}}}]{Treu2002}
\bibinfo{author}{\bibfnamefont{T.}~\bibnamefont{{Treu}}},
  \bibinfo{author}{\bibfnamefont{M.}~\bibnamefont{{Stiavelli}}},
  \bibinfo{author}{\bibfnamefont{S.}~\bibnamefont{{Casertano}}},
  \bibinfo{author}{\bibfnamefont{P.}~\bibnamefont{{M{\o}ller}}},
  \bibnamefont{and} \bibinfo{author}{\bibfnamefont{G.}~\bibnamefont{{Bertin}}},
  \bibinfo{journal}{Astrophysical Journal l} \textbf{\bibinfo{volume}{564}},
  \bibinfo{pages}{L13} (\bibinfo{year}{2002}), \eprint{astro-ph/0111504}.

\bibitem[{\citenamefont{{Stern}
  et~al.}(2010{\natexlab{a}})\citenamefont{{Stern}, {Jimenez}, {Verde},
  {Kamionkowski}, and {Stanford}}}]{stern10}
\bibinfo{author}{\bibfnamefont{D.}~\bibnamefont{{Stern}}},
  \bibinfo{author}{\bibfnamefont{R.}~\bibnamefont{{Jimenez}}},
  \bibinfo{author}{\bibfnamefont{L.}~\bibnamefont{{Verde}}},
  \bibinfo{author}{\bibfnamefont{M.}~\bibnamefont{{Kamionkowski}}},
  \bibnamefont{and} \bibinfo{author}{\bibfnamefont{S.~A.}
  \bibnamefont{{Stanford}}}, \bibinfo{journal}{\jcap}
  \textbf{\bibinfo{volume}{2}}, \bibinfo{eid}{008}
  (\bibinfo{year}{2010}{\natexlab{a}}), \eprint{0907.3149}.

\bibitem[{\citenamefont{{Stern} et~al.}(2001)\citenamefont{{Stern}, {Connolly},
  {Eisenhardt}, {Elston}, {Holden}, {Rosati}, {Stanford}, {Spinrad}, {Tozzi},
  and {Wu}}}]{Stern2001}
\bibinfo{author}{\bibfnamefont{D.}~\bibnamefont{{Stern}}},
  \bibinfo{author}{\bibfnamefont{A.}~\bibnamefont{{Connolly}}},
  \bibinfo{author}{\bibfnamefont{P.}~\bibnamefont{{Eisenhardt}}},
  \bibinfo{author}{\bibfnamefont{R.}~\bibnamefont{{Elston}}},
  \bibinfo{author}{\bibfnamefont{B.}~\bibnamefont{{Holden}}},
  \bibinfo{author}{\bibfnamefont{P.}~\bibnamefont{{Rosati}}},
  \bibinfo{author}{\bibfnamefont{S.~A.} \bibnamefont{{Stanford}}},
  \bibinfo{author}{\bibfnamefont{H.}~\bibnamefont{{Spinrad}}},
  \bibinfo{author}{\bibfnamefont{P.}~\bibnamefont{{Tozzi}}}, \bibnamefont{and}
  \bibinfo{author}{\bibfnamefont{K.}~\bibnamefont{{Wu}}}, in
  \emph{\bibinfo{booktitle}{Deep Fields}}, edited by
  \bibinfo{editor}{\bibfnamefont{S.}~\bibnamefont{{Cristiani}}},
  \bibinfo{editor}{\bibfnamefont{A.}~\bibnamefont{{Renzini}}},
  \bibnamefont{and} \bibinfo{editor}{\bibfnamefont{R.~E.}
  \bibnamefont{{Williams}}} (\bibinfo{year}{2001}), p.~\bibinfo{pages}{76},
  \eprint{astro-ph/0012146}.

\bibitem[{\citenamefont{{Le F{\`e}vre} et~al.}(2005)\citenamefont{{Le
  F{\`e}vre}, {Vettolani}, {Garilli}, {Tresse}, {Bottini}, {Le Brun},
  {Maccagni}, {Picat}, {Scaramella}, {Scodeggio} et~al.}}]{LeFevre2005}
\bibinfo{author}{\bibfnamefont{O.}~\bibnamefont{{Le F{\`e}vre}}},
  \bibinfo{author}{\bibfnamefont{G.}~\bibnamefont{{Vettolani}}},
  \bibinfo{author}{\bibfnamefont{B.}~\bibnamefont{{Garilli}}},
  \bibinfo{author}{\bibfnamefont{L.}~\bibnamefont{{Tresse}}},
  \bibinfo{author}{\bibfnamefont{D.}~\bibnamefont{{Bottini}}},
  \bibinfo{author}{\bibfnamefont{V.}~\bibnamefont{{Le Brun}}},
  \bibinfo{author}{\bibfnamefont{D.}~\bibnamefont{{Maccagni}}},
  \bibinfo{author}{\bibfnamefont{J.~P.} \bibnamefont{{Picat}}},
  \bibinfo{author}{\bibfnamefont{R.}~\bibnamefont{{Scaramella}}},
  \bibinfo{author}{\bibfnamefont{M.}~\bibnamefont{{Scodeggio}}},
  \bibnamefont{et~al.}, \bibinfo{journal}{Astronomy \& Astrophysics}
  \textbf{\bibinfo{volume}{439}}, \bibinfo{pages}{845} (\bibinfo{year}{2005}),
  \eprint{astro-ph/0409133}.

\bibitem[{\citenamefont{{Moresco} et~al.}(2012)\citenamefont{{Moresco},
  {Cimatti}, {Jimenez}, {Pozzetti}, {Zamorani}, {Bolzonella}, {Dunlop},
  {Lamareille}, {Mignoli}, {Pearce} et~al.}}]{moresco12}
\bibinfo{author}{\bibfnamefont{M.}~\bibnamefont{{Moresco}}},
  \bibinfo{author}{\bibfnamefont{A.}~\bibnamefont{{Cimatti}}},
  \bibinfo{author}{\bibfnamefont{R.}~\bibnamefont{{Jimenez}}},
  \bibinfo{author}{\bibfnamefont{L.}~\bibnamefont{{Pozzetti}}},
  \bibinfo{author}{\bibfnamefont{G.}~\bibnamefont{{Zamorani}}},
  \bibinfo{author}{\bibfnamefont{M.}~\bibnamefont{{Bolzonella}}},
  \bibinfo{author}{\bibfnamefont{J.}~\bibnamefont{{Dunlop}}},
  \bibinfo{author}{\bibfnamefont{F.}~\bibnamefont{{Lamareille}}},
  \bibinfo{author}{\bibfnamefont{M.}~\bibnamefont{{Mignoli}}},
  \bibinfo{author}{\bibfnamefont{H.}~\bibnamefont{{Pearce}}},
  \bibnamefont{et~al.}, \bibinfo{journal}{\jcap} \textbf{\bibinfo{volume}{8}},
  \bibinfo{eid}{006} (\bibinfo{year}{2012}), \eprint{1201.3609}.

\bibitem[{\citenamefont{{Cimatti} et~al.}(2002)\citenamefont{{Cimatti},
  {Pozzetti}, {Mignoli}, {Daddi}, {Menci}, {Poli}, {Fontana}, {Renzini},
  {Zamorani}, {Broadhurst} et~al.}}]{Cimatti2002}
\bibinfo{author}{\bibfnamefont{A.}~\bibnamefont{{Cimatti}}},
  \bibinfo{author}{\bibfnamefont{L.}~\bibnamefont{{Pozzetti}}},
  \bibinfo{author}{\bibfnamefont{M.}~\bibnamefont{{Mignoli}}},
  \bibinfo{author}{\bibfnamefont{E.}~\bibnamefont{{Daddi}}},
  \bibinfo{author}{\bibfnamefont{N.}~\bibnamefont{{Menci}}},
  \bibinfo{author}{\bibfnamefont{F.}~\bibnamefont{{Poli}}},
  \bibinfo{author}{\bibfnamefont{A.}~\bibnamefont{{Fontana}}},
  \bibinfo{author}{\bibfnamefont{A.}~\bibnamefont{{Renzini}}},
  \bibinfo{author}{\bibfnamefont{G.}~\bibnamefont{{Zamorani}}},
  \bibinfo{author}{\bibfnamefont{T.}~\bibnamefont{{Broadhurst}}},
  \bibnamefont{et~al.}, \bibinfo{journal}{Astronomy \& Astrophysics}
  \textbf{\bibinfo{volume}{391}}, \bibinfo{pages}{L1} (\bibinfo{year}{2002}),
  \eprint{astro-ph/0207191}.

\bibitem[{\citenamefont{{Demarco} et~al.}(2010)\citenamefont{{Demarco},
  {Gobat}, {Rosati}, {Lidman}, {Rettura}, {Nonino}, {van der Wel}, {Jee},
  {Blakeslee}, {Ford} et~al.}}]{Demarco2010}
\bibinfo{author}{\bibfnamefont{R.}~\bibnamefont{{Demarco}}},
  \bibinfo{author}{\bibfnamefont{R.}~\bibnamefont{{Gobat}}},
  \bibinfo{author}{\bibfnamefont{P.}~\bibnamefont{{Rosati}}},
  \bibinfo{author}{\bibfnamefont{C.}~\bibnamefont{{Lidman}}},
  \bibinfo{author}{\bibfnamefont{A.}~\bibnamefont{{Rettura}}},
  \bibinfo{author}{\bibfnamefont{M.}~\bibnamefont{{Nonino}}},
  \bibinfo{author}{\bibfnamefont{A.}~\bibnamefont{{van der Wel}}},
  \bibinfo{author}{\bibfnamefont{M.~J.} \bibnamefont{{Jee}}},
  \bibinfo{author}{\bibfnamefont{J.~P.} \bibnamefont{{Blakeslee}}},
  \bibinfo{author}{\bibfnamefont{H.~C.} \bibnamefont{{Ford}}},
  \bibnamefont{et~al.}, \bibinfo{journal}{Astrophysical Journal}
  \textbf{\bibinfo{volume}{725}}, \bibinfo{pages}{1252} (\bibinfo{year}{2010}),
  \eprint{1009.3986}.

\bibitem[{\citenamefont{{Eisenstein} et~al.}(2001)\citenamefont{{Eisenstein},
  {Annis}, {Gunn}, {Szalay}, {Connolly}, {Nichol}, {Bahcall}, {Bernardi},
  {Burles}, {Castander} et~al.}}]{Eisenstein2001}
\bibinfo{author}{\bibfnamefont{D.~J.} \bibnamefont{{Eisenstein}}},
  \bibinfo{author}{\bibfnamefont{J.}~\bibnamefont{{Annis}}},
  \bibinfo{author}{\bibfnamefont{J.~E.} \bibnamefont{{Gunn}}},
  \bibinfo{author}{\bibfnamefont{A.~S.} \bibnamefont{{Szalay}}},
  \bibinfo{author}{\bibfnamefont{A.~J.} \bibnamefont{{Connolly}}},
  \bibinfo{author}{\bibfnamefont{R.~C.} \bibnamefont{{Nichol}}},
  \bibinfo{author}{\bibfnamefont{N.~A.} \bibnamefont{{Bahcall}}},
  \bibinfo{author}{\bibfnamefont{M.}~\bibnamefont{{Bernardi}}},
  \bibinfo{author}{\bibfnamefont{S.}~\bibnamefont{{Burles}}},
  \bibinfo{author}{\bibfnamefont{F.~J.} \bibnamefont{{Castander}}},
  \bibnamefont{et~al.}, \bibinfo{journal}{Astronomical Journal}
  \textbf{\bibinfo{volume}{122}}, \bibinfo{pages}{2267} (\bibinfo{year}{2001}),
  \eprint{astro-ph/0108153}.

\bibitem[{\citenamefont{{Le Borgne} et~al.}(2006)\citenamefont{{Le Borgne},
  {Abraham}, {Daniel}, {McCarthy}, {Glazebrook}, {Savaglio}, {Crampton},
  {Juneau}, {Carlberg}, {Chen} et~al.}}]{LeBorgne2006}
\bibinfo{author}{\bibfnamefont{D.}~\bibnamefont{{Le Borgne}}},
  \bibinfo{author}{\bibfnamefont{R.}~\bibnamefont{{Abraham}}},
  \bibinfo{author}{\bibfnamefont{K.}~\bibnamefont{{Daniel}}},
  \bibinfo{author}{\bibfnamefont{P.~J.} \bibnamefont{{McCarthy}}},
  \bibinfo{author}{\bibfnamefont{K.}~\bibnamefont{{Glazebrook}}},
  \bibinfo{author}{\bibfnamefont{S.}~\bibnamefont{{Savaglio}}},
  \bibinfo{author}{\bibfnamefont{D.}~\bibnamefont{{Crampton}}},
  \bibinfo{author}{\bibfnamefont{S.}~\bibnamefont{{Juneau}}},
  \bibinfo{author}{\bibfnamefont{R.~G.} \bibnamefont{{Carlberg}}},
  \bibinfo{author}{\bibfnamefont{H.-W.} \bibnamefont{{Chen}}},
  \bibnamefont{et~al.}, \bibinfo{journal}{Astrophysical Journal}
  \textbf{\bibinfo{volume}{642}}, \bibinfo{pages}{48} (\bibinfo{year}{2006}),
  \eprint{astro-ph/0503401}.

\bibitem[{\citenamefont{{Lilly} et~al.}(2009)\citenamefont{{Lilly}, {Le Brun},
  {Maier}, {Mainieri}, {Mignoli}, {Scodeggio}, {Zamorani}, {Carollo},
  {Contini}, {Kneib} et~al.}}]{Lilly2009}
\bibinfo{author}{\bibfnamefont{S.~J.} \bibnamefont{{Lilly}}},
  \bibinfo{author}{\bibfnamefont{V.}~\bibnamefont{{Le Brun}}},
  \bibinfo{author}{\bibfnamefont{C.}~\bibnamefont{{Maier}}},
  \bibinfo{author}{\bibfnamefont{V.}~\bibnamefont{{Mainieri}}},
  \bibinfo{author}{\bibfnamefont{M.}~\bibnamefont{{Mignoli}}},
  \bibinfo{author}{\bibfnamefont{M.}~\bibnamefont{{Scodeggio}}},
  \bibinfo{author}{\bibfnamefont{G.}~\bibnamefont{{Zamorani}}},
  \bibinfo{author}{\bibfnamefont{M.}~\bibnamefont{{Carollo}}},
  \bibinfo{author}{\bibfnamefont{T.}~\bibnamefont{{Contini}}},
  \bibinfo{author}{\bibfnamefont{J.-P.} \bibnamefont{{Kneib}}},
  \bibnamefont{et~al.}, \bibinfo{journal}{Astrophysical Journal Supplement
  Series} \textbf{\bibinfo{volume}{184}}, \bibinfo{pages}{218}
  (\bibinfo{year}{2009}).

\bibitem[{\citenamefont{{Onodera} et~al.}(2010)\citenamefont{{Onodera},
  {Daddi}, {Gobat}, {Cappellari}, {Arimoto}, {Renzini}, {Yamada}, {McCracken},
  {Mancini}, {Capak} et~al.}}]{Onodera2010}
\bibinfo{author}{\bibfnamefont{M.}~\bibnamefont{{Onodera}}},
  \bibinfo{author}{\bibfnamefont{E.}~\bibnamefont{{Daddi}}},
  \bibinfo{author}{\bibfnamefont{R.}~\bibnamefont{{Gobat}}},
  \bibinfo{author}{\bibfnamefont{M.}~\bibnamefont{{Cappellari}}},
  \bibinfo{author}{\bibfnamefont{N.}~\bibnamefont{{Arimoto}}},
  \bibinfo{author}{\bibfnamefont{A.}~\bibnamefont{{Renzini}}},
  \bibinfo{author}{\bibfnamefont{Y.}~\bibnamefont{{Yamada}}},
  \bibinfo{author}{\bibfnamefont{H.~J.} \bibnamefont{{McCracken}}},
  \bibinfo{author}{\bibfnamefont{C.}~\bibnamefont{{Mancini}}},
  \bibinfo{author}{\bibfnamefont{P.}~\bibnamefont{{Capak}}},
  \bibnamefont{et~al.}, \bibinfo{journal}{Astrophysical Journal l}
  \textbf{\bibinfo{volume}{715}}, \bibinfo{pages}{L6} (\bibinfo{year}{2010}),
  \eprint{1004.2120}.

\bibitem[{\citenamefont{{Rosati} et~al.}(2009)\citenamefont{{Rosati}, {Tozzi},
  {Gobat}, {Santos}, {Nonino}, {Demarco}, {Lidman}, {Mullis}, {Strazzullo},
  {B{\"o}hringer} et~al.}}]{Rosati2009}
\bibinfo{author}{\bibfnamefont{P.}~\bibnamefont{{Rosati}}},
  \bibinfo{author}{\bibfnamefont{P.}~\bibnamefont{{Tozzi}}},
  \bibinfo{author}{\bibfnamefont{R.}~\bibnamefont{{Gobat}}},
  \bibinfo{author}{\bibfnamefont{J.~S.} \bibnamefont{{Santos}}},
  \bibinfo{author}{\bibfnamefont{M.}~\bibnamefont{{Nonino}}},
  \bibinfo{author}{\bibfnamefont{R.}~\bibnamefont{{Demarco}}},
  \bibinfo{author}{\bibfnamefont{C.}~\bibnamefont{{Lidman}}},
  \bibinfo{author}{\bibfnamefont{C.~R.} \bibnamefont{{Mullis}}},
  \bibinfo{author}{\bibfnamefont{V.}~\bibnamefont{{Strazzullo}}},
  \bibinfo{author}{\bibfnamefont{H.}~\bibnamefont{{B{\"o}hringer}}},
  \bibnamefont{et~al.}, \bibinfo{journal}{Astronomy \& Astrophysics}
  \textbf{\bibinfo{volume}{508}}, \bibinfo{pages}{583} (\bibinfo{year}{2009}),
  \eprint{0910.1716}.

\bibitem[{\citenamefont{{Stern}
  et~al.}(2010{\natexlab{b}})\citenamefont{{Stern}, {Jimenez}, {Verde},
  {Kamionkowski}, and {Stanford}}}]{Stern2010}
\bibinfo{author}{\bibfnamefont{D.}~\bibnamefont{{Stern}}},
  \bibinfo{author}{\bibfnamefont{R.}~\bibnamefont{{Jimenez}}},
  \bibinfo{author}{\bibfnamefont{L.}~\bibnamefont{{Verde}}},
  \bibinfo{author}{\bibfnamefont{M.}~\bibnamefont{{Kamionkowski}}},
  \bibnamefont{and} \bibinfo{author}{\bibfnamefont{S.~A.}
  \bibnamefont{{Stanford}}}, \bibinfo{journal}{\jcap}
  \textbf{\bibinfo{volume}{2010}}, \bibinfo{eid}{008}
  (\bibinfo{year}{2010}{\natexlab{b}}), \eprint{0907.3149}.

\bibitem[{\citenamefont{{Strauss} et~al.}(2002)\citenamefont{{Strauss},
  {Weinberg}, {Lupton}, {Narayanan}, {Annis}, {Bernardi}, {Blanton}, {Burles},
  {Connolly}, {Dalcanton} et~al.}}]{Strauss2002}
\bibinfo{author}{\bibfnamefont{M.~A.} \bibnamefont{{Strauss}}},
  \bibinfo{author}{\bibfnamefont{D.~H.} \bibnamefont{{Weinberg}}},
  \bibinfo{author}{\bibfnamefont{R.~H.} \bibnamefont{{Lupton}}},
  \bibinfo{author}{\bibfnamefont{V.~K.} \bibnamefont{{Narayanan}}},
  \bibinfo{author}{\bibfnamefont{J.}~\bibnamefont{{Annis}}},
  \bibinfo{author}{\bibfnamefont{M.}~\bibnamefont{{Bernardi}}},
  \bibinfo{author}{\bibfnamefont{M.}~\bibnamefont{{Blanton}}},
  \bibinfo{author}{\bibfnamefont{S.}~\bibnamefont{{Burles}}},
  \bibinfo{author}{\bibfnamefont{A.~J.} \bibnamefont{{Connolly}}},
  \bibinfo{author}{\bibfnamefont{J.}~\bibnamefont{{Dalcanton}}},
  \bibnamefont{et~al.}, \bibinfo{journal}{Astronomical Journal}
  \textbf{\bibinfo{volume}{124}}, \bibinfo{pages}{1810} (\bibinfo{year}{2002}),
  \eprint{astro-ph/0206225}.

\bibitem[{\citenamefont{{Vanzella} et~al.}(2008)\citenamefont{{Vanzella},
  {Cristiani}, {Dickinson}, {Giavalisco}, {Kuntschner}, {Haase}, {Nonino},
  {Rosati}, {Cesarsky}, {Ferguson} et~al.}}]{Vanzella2008}
\bibinfo{author}{\bibfnamefont{E.}~\bibnamefont{{Vanzella}}},
  \bibinfo{author}{\bibfnamefont{S.}~\bibnamefont{{Cristiani}}},
  \bibinfo{author}{\bibfnamefont{M.}~\bibnamefont{{Dickinson}}},
  \bibinfo{author}{\bibfnamefont{M.}~\bibnamefont{{Giavalisco}}},
  \bibinfo{author}{\bibfnamefont{H.}~\bibnamefont{{Kuntschner}}},
  \bibinfo{author}{\bibfnamefont{J.}~\bibnamefont{{Haase}}},
  \bibinfo{author}{\bibfnamefont{M.}~\bibnamefont{{Nonino}}},
  \bibinfo{author}{\bibfnamefont{P.}~\bibnamefont{{Rosati}}},
  \bibinfo{author}{\bibfnamefont{C.}~\bibnamefont{{Cesarsky}}},
  \bibinfo{author}{\bibfnamefont{H.~C.} \bibnamefont{{Ferguson}}},
  \bibnamefont{et~al.}, \bibinfo{journal}{Astronomy \& Astrophysics}
  \textbf{\bibinfo{volume}{478}}, \bibinfo{pages}{83} (\bibinfo{year}{2008}),
  \eprint{0711.0850}.

\bibitem[{\citenamefont{{Zhang} et~al.}(2014)\citenamefont{{Zhang}, {Zhang},
  {Yuan}, {Liu}, {Zhang}, and {Sun}}}]{zhang14}
\bibinfo{author}{\bibfnamefont{C.}~\bibnamefont{{Zhang}}},
  \bibinfo{author}{\bibfnamefont{H.}~\bibnamefont{{Zhang}}},
  \bibinfo{author}{\bibfnamefont{S.}~\bibnamefont{{Yuan}}},
  \bibinfo{author}{\bibfnamefont{S.}~\bibnamefont{{Liu}}},
  \bibinfo{author}{\bibfnamefont{T.-J.} \bibnamefont{{Zhang}}},
  \bibnamefont{and} \bibinfo{author}{\bibfnamefont{Y.-C.} \bibnamefont{{Sun}}},
  \bibinfo{journal}{Research in Astronomy and Astrophysics}
  \textbf{\bibinfo{volume}{14}}, \bibinfo{eid}{1221-1233}
  (\bibinfo{year}{2014}), \eprint{1207.4541}.

\bibitem[{\citenamefont{{Abazajian} et~al.}(2009)\citenamefont{{Abazajian},
  {Adelman-McCarthy}, {Ag{\"u}eros}, {Allam}, {Allende Prieto}, {An},
  {Anderson}, {Anderson}, {Annis}, {Bahcall} et~al.}}]{Abazajian2009}
\bibinfo{author}{\bibfnamefont{K.~N.} \bibnamefont{{Abazajian}}},
  \bibinfo{author}{\bibfnamefont{J.~K.} \bibnamefont{{Adelman-McCarthy}}},
  \bibinfo{author}{\bibfnamefont{M.~A.} \bibnamefont{{Ag{\"u}eros}}},
  \bibinfo{author}{\bibfnamefont{S.~S.} \bibnamefont{{Allam}}},
  \bibinfo{author}{\bibfnamefont{C.}~\bibnamefont{{Allende Prieto}}},
  \bibinfo{author}{\bibfnamefont{D.}~\bibnamefont{{An}}},
  \bibinfo{author}{\bibfnamefont{K.~S.~J.} \bibnamefont{{Anderson}}},
  \bibinfo{author}{\bibfnamefont{S.~F.} \bibnamefont{{Anderson}}},
  \bibinfo{author}{\bibfnamefont{J.}~\bibnamefont{{Annis}}},
  \bibinfo{author}{\bibfnamefont{N.~A.} \bibnamefont{{Bahcall}}},
  \bibnamefont{et~al.}, \bibinfo{journal}{Astrophysical Journal Supplement
  Series} \textbf{\bibinfo{volume}{182}}, \bibinfo{pages}{543}
  (\bibinfo{year}{2009}), \eprint{0812.0649}.

\bibitem[{\citenamefont{{Moresco}}(2015)}]{moresco15}
\bibinfo{author}{\bibfnamefont{M.}~\bibnamefont{{Moresco}}},
  \bibinfo{journal}{\mnras} \textbf{\bibinfo{volume}{450}},
  \bibinfo{pages}{L16} (\bibinfo{year}{2015}), \eprint{1503.01116}.

\bibitem[{\citenamefont{{Gobat} et~al.}(2013)\citenamefont{{Gobat},
  {Strazzullo}, {Daddi}, {Onodera}, {Carollo}, {Renzini}, {Finoguenov},
  {Cimatti}, {Scarlata}, and {Arimoto}}}]{Gobat2013}
\bibinfo{author}{\bibfnamefont{R.}~\bibnamefont{{Gobat}}},
  \bibinfo{author}{\bibfnamefont{V.}~\bibnamefont{{Strazzullo}}},
  \bibinfo{author}{\bibfnamefont{E.}~\bibnamefont{{Daddi}}},
  \bibinfo{author}{\bibfnamefont{M.}~\bibnamefont{{Onodera}}},
  \bibinfo{author}{\bibfnamefont{M.}~\bibnamefont{{Carollo}}},
  \bibinfo{author}{\bibfnamefont{A.}~\bibnamefont{{Renzini}}},
  \bibinfo{author}{\bibfnamefont{A.}~\bibnamefont{{Finoguenov}}},
  \bibinfo{author}{\bibfnamefont{A.}~\bibnamefont{{Cimatti}}},
  \bibinfo{author}{\bibfnamefont{C.}~\bibnamefont{{Scarlata}}},
  \bibnamefont{and}
  \bibinfo{author}{\bibfnamefont{N.}~\bibnamefont{{Arimoto}}},
  \bibinfo{journal}{Astrophysical Journal} \textbf{\bibinfo{volume}{776}},
  \bibinfo{eid}{9} (\bibinfo{year}{2013}), \eprint{1305.3576}.

\bibitem[{\citenamefont{{Kriek} et~al.}(2009)\citenamefont{{Kriek}, {van
  Dokkum}, {Labb{\'e}}, {Franx}, {Illingworth}, {Marchesini}, and
  {Quadri}}}]{Kriek2009}
\bibinfo{author}{\bibfnamefont{M.}~\bibnamefont{{Kriek}}},
  \bibinfo{author}{\bibfnamefont{P.~G.} \bibnamefont{{van Dokkum}}},
  \bibinfo{author}{\bibfnamefont{I.}~\bibnamefont{{Labb{\'e}}}},
  \bibinfo{author}{\bibfnamefont{M.}~\bibnamefont{{Franx}}},
  \bibinfo{author}{\bibfnamefont{G.~D.} \bibnamefont{{Illingworth}}},
  \bibinfo{author}{\bibfnamefont{D.}~\bibnamefont{{Marchesini}}},
  \bibnamefont{and} \bibinfo{author}{\bibfnamefont{R.~F.}
  \bibnamefont{{Quadri}}}, \bibinfo{journal}{Astrophysical Journal}
  \textbf{\bibinfo{volume}{700}}, \bibinfo{pages}{221} (\bibinfo{year}{2009}),
  \eprint{0905.1692}.

\bibitem[{\citenamefont{{Krogager} et~al.}(2014)\citenamefont{{Krogager},
  {Zirm}, {Toft}, {Man}, and {Brammer}}}]{Krogager2014}
\bibinfo{author}{\bibfnamefont{J.~K.} \bibnamefont{{Krogager}}},
  \bibinfo{author}{\bibfnamefont{A.~W.} \bibnamefont{{Zirm}}},
  \bibinfo{author}{\bibfnamefont{S.}~\bibnamefont{{Toft}}},
  \bibinfo{author}{\bibfnamefont{A.}~\bibnamefont{{Man}}}, \bibnamefont{and}
  \bibinfo{author}{\bibfnamefont{G.}~\bibnamefont{{Brammer}}},
  \bibinfo{journal}{Astrophysical Journal} \textbf{\bibinfo{volume}{797}},
  \bibinfo{eid}{17} (\bibinfo{year}{2014}), \eprint{1309.6316}.

\bibitem[{\citenamefont{{Onodera} et~al.}(2012)\citenamefont{{Onodera},
  {Renzini}, {Carollo}, {Cappellari}, {Mancini}, {Strazzullo}, {Daddi},
  {Arimoto}, {Gobat}, {Yamada} et~al.}}]{Onodera2012}
\bibinfo{author}{\bibfnamefont{M.}~\bibnamefont{{Onodera}}},
  \bibinfo{author}{\bibfnamefont{A.}~\bibnamefont{{Renzini}}},
  \bibinfo{author}{\bibfnamefont{M.}~\bibnamefont{{Carollo}}},
  \bibinfo{author}{\bibfnamefont{M.}~\bibnamefont{{Cappellari}}},
  \bibinfo{author}{\bibfnamefont{C.}~\bibnamefont{{Mancini}}},
  \bibinfo{author}{\bibfnamefont{V.}~\bibnamefont{{Strazzullo}}},
  \bibinfo{author}{\bibfnamefont{E.}~\bibnamefont{{Daddi}}},
  \bibinfo{author}{\bibfnamefont{N.}~\bibnamefont{{Arimoto}}},
  \bibinfo{author}{\bibfnamefont{R.}~\bibnamefont{{Gobat}}},
  \bibinfo{author}{\bibfnamefont{Y.}~\bibnamefont{{Yamada}}},
  \bibnamefont{et~al.}, \bibinfo{journal}{Astrophysical Journal}
  \textbf{\bibinfo{volume}{755}}, \bibinfo{eid}{26} (\bibinfo{year}{2012}),
  \eprint{1206.1540}.

\bibitem[{\citenamefont{{Saracco} et~al.}(2005)\citenamefont{{Saracco},
  {Longhetti}, {Severgnini}, {Della Ceca}, {Braito}, {Mannucci}, {Bender},
  {Drory}, {Feulner}, {Hopp} et~al.}}]{Saracco2005}
\bibinfo{author}{\bibfnamefont{P.}~\bibnamefont{{Saracco}}},
  \bibinfo{author}{\bibfnamefont{M.}~\bibnamefont{{Longhetti}}},
  \bibinfo{author}{\bibfnamefont{P.}~\bibnamefont{{Severgnini}}},
  \bibinfo{author}{\bibfnamefont{R.}~\bibnamefont{{Della Ceca}}},
  \bibinfo{author}{\bibfnamefont{V.}~\bibnamefont{{Braito}}},
  \bibinfo{author}{\bibfnamefont{F.}~\bibnamefont{{Mannucci}}},
  \bibinfo{author}{\bibfnamefont{R.}~\bibnamefont{{Bender}}},
  \bibinfo{author}{\bibfnamefont{N.}~\bibnamefont{{Drory}}},
  \bibinfo{author}{\bibfnamefont{G.}~\bibnamefont{{Feulner}}},
  \bibinfo{author}{\bibfnamefont{U.}~\bibnamefont{{Hopp}}},
  \bibnamefont{et~al.}, \bibinfo{journal}{\mnras}
  \textbf{\bibinfo{volume}{357}}, \bibinfo{pages}{L40} (\bibinfo{year}{2005}),
  \eprint{astro-ph/0412020}.

\bibitem[{\citenamefont{{Moresco} et~al.}(2016)\citenamefont{{Moresco},
  {Pozzetti}, {Cimatti}, {Jimenez}, {Maraston}, {Verde}, {Thomas}, {Citro},
  {Tojeiro}, and {Wilkinson}}}]{CC2}
\bibinfo{author}{\bibfnamefont{M.}~\bibnamefont{{Moresco}}},
  \bibinfo{author}{\bibfnamefont{L.}~\bibnamefont{{Pozzetti}}},
  \bibinfo{author}{\bibfnamefont{A.}~\bibnamefont{{Cimatti}}},
  \bibinfo{author}{\bibfnamefont{R.}~\bibnamefont{{Jimenez}}},
  \bibinfo{author}{\bibfnamefont{C.}~\bibnamefont{{Maraston}}},
  \bibinfo{author}{\bibfnamefont{L.}~\bibnamefont{{Verde}}},
  \bibinfo{author}{\bibfnamefont{D.}~\bibnamefont{{Thomas}}},
  \bibinfo{author}{\bibfnamefont{A.}~\bibnamefont{{Citro}}},
  \bibinfo{author}{\bibfnamefont{R.}~\bibnamefont{{Tojeiro}}},
  \bibnamefont{and}
  \bibinfo{author}{\bibfnamefont{D.}~\bibnamefont{{Wilkinson}}},
  \bibinfo{journal}{\jcap} \textbf{\bibinfo{volume}{5}}, \bibinfo{eid}{014}
  (\bibinfo{year}{2016}), \eprint{1601.01701}.

\bibitem[{\citenamefont{{Dawson} et~al.}(2013)\citenamefont{{Dawson},
  {Schlegel}, {Ahn}, {Anderson}, {Aubourg}, {Bailey}, {Barkhouser}, {Bautista},
  {Beifiori}, {Berlind} et~al.}}]{Dawson2013}
\bibinfo{author}{\bibfnamefont{K.~S.} \bibnamefont{{Dawson}}},
  \bibinfo{author}{\bibfnamefont{D.~J.} \bibnamefont{{Schlegel}}},
  \bibinfo{author}{\bibfnamefont{C.~P.} \bibnamefont{{Ahn}}},
  \bibinfo{author}{\bibfnamefont{S.~F.} \bibnamefont{{Anderson}}},
  \bibinfo{author}{\bibfnamefont{{\'E}.}~\bibnamefont{{Aubourg}}},
  \bibinfo{author}{\bibfnamefont{S.}~\bibnamefont{{Bailey}}},
  \bibinfo{author}{\bibfnamefont{R.~H.} \bibnamefont{{Barkhouser}}},
  \bibinfo{author}{\bibfnamefont{J.~E.} \bibnamefont{{Bautista}}},
  \bibinfo{author}{\bibfnamefont{A.~r.} \bibnamefont{{Beifiori}}},
  \bibinfo{author}{\bibfnamefont{A.~A.} \bibnamefont{{Berlind}}},
  \bibnamefont{et~al.}, \bibinfo{journal}{Astronomical Journal}
  \textbf{\bibinfo{volume}{145}}, \bibinfo{eid}{10} (\bibinfo{year}{2013}),
  \eprint{1208.0022}.

\bibitem[{\citenamefont{{Eisenstein} et~al.}(2011)\citenamefont{{Eisenstein},
  {Weinberg}, {Agol}, {Aihara}, {Allende Prieto}, {Anderson}, {Arns},
  {Aubourg}, {Bailey}, {Balbinot} et~al.}}]{Eisenstein2011}
\bibinfo{author}{\bibfnamefont{D.~J.} \bibnamefont{{Eisenstein}}},
  \bibinfo{author}{\bibfnamefont{D.~H.} \bibnamefont{{Weinberg}}},
  \bibinfo{author}{\bibfnamefont{E.}~\bibnamefont{{Agol}}},
  \bibinfo{author}{\bibfnamefont{H.}~\bibnamefont{{Aihara}}},
  \bibinfo{author}{\bibfnamefont{C.}~\bibnamefont{{Allende Prieto}}},
  \bibinfo{author}{\bibfnamefont{S.~F.} \bibnamefont{{Anderson}}},
  \bibinfo{author}{\bibfnamefont{J.~A.} \bibnamefont{{Arns}}},
  \bibinfo{author}{\bibfnamefont{{\'E}.}~\bibnamefont{{Aubourg}}},
  \bibinfo{author}{\bibfnamefont{S.}~\bibnamefont{{Bailey}}},
  \bibinfo{author}{\bibfnamefont{E.}~\bibnamefont{{Balbinot}}},
  \bibnamefont{et~al.}, \bibinfo{journal}{Astronomical Journal}
  \textbf{\bibinfo{volume}{142}}, \bibinfo{eid}{72} (\bibinfo{year}{2011}),
  \eprint{1101.1529}.

\bibitem[{\citenamefont{{Beutler} et~al.}(2011)\citenamefont{{Beutler},
  {Blake}, {Colless}, {Jones}, {Staveley-Smith}, {Campbell}, {Parker},
  {Saunders}, and {Watson}}}]{6dFGS}
\bibinfo{author}{\bibfnamefont{F.}~\bibnamefont{{Beutler}}},
  \bibinfo{author}{\bibfnamefont{C.}~\bibnamefont{{Blake}}},
  \bibinfo{author}{\bibfnamefont{M.}~\bibnamefont{{Colless}}},
  \bibinfo{author}{\bibfnamefont{D.~H.} \bibnamefont{{Jones}}},
  \bibinfo{author}{\bibfnamefont{L.}~\bibnamefont{{Staveley-Smith}}},
  \bibinfo{author}{\bibfnamefont{L.}~\bibnamefont{{Campbell}}},
  \bibinfo{author}{\bibfnamefont{Q.}~\bibnamefont{{Parker}}},
  \bibinfo{author}{\bibfnamefont{W.}~\bibnamefont{{Saunders}}},
  \bibnamefont{and} \bibinfo{author}{\bibfnamefont{F.}~\bibnamefont{{Watson}}},
  \bibinfo{journal}{\mnras} \textbf{\bibinfo{volume}{416}},
  \bibinfo{pages}{3017} (\bibinfo{year}{2011}), \eprint{1106.3366}.

\bibitem[{\citenamefont{Kazin et~al.}(2014)\citenamefont{Kazin, Koda, Blake,
  Padmanabhan, Brough, Colless, Contreras, Couch, Croom, Croton
  et~al.}}]{WiggleZ}
\bibinfo{author}{\bibfnamefont{E.~A.} \bibnamefont{Kazin}},
  \bibinfo{author}{\bibfnamefont{J.}~\bibnamefont{Koda}},
  \bibinfo{author}{\bibfnamefont{C.}~\bibnamefont{Blake}},
  \bibinfo{author}{\bibfnamefont{N.}~\bibnamefont{Padmanabhan}},
  \bibinfo{author}{\bibfnamefont{S.}~\bibnamefont{Brough}},
  \bibinfo{author}{\bibfnamefont{M.}~\bibnamefont{Colless}},
  \bibinfo{author}{\bibfnamefont{C.}~\bibnamefont{Contreras}},
  \bibinfo{author}{\bibfnamefont{W.}~\bibnamefont{Couch}},
  \bibinfo{author}{\bibfnamefont{S.}~\bibnamefont{Croom}},
  \bibinfo{author}{\bibfnamefont{D.~J.} \bibnamefont{Croton}},
  \bibnamefont{et~al.}, \bibinfo{journal}{Monthly Notices of the Royal
  Astronomical Society} \textbf{\bibinfo{volume}{441}},
  \bibinfo{pages}{3524–3542} (\bibinfo{year}{2014}), ISSN
  \bibinfo{issn}{0035-8711},
  \urlprefix\url{http://dx.doi.org/10.1093/mnras/stu778}.

\bibitem[{\citenamefont{Ata et~al.}(2017)\citenamefont{Ata, Baumgarten,
  Bautista, Beutler, Bizyaev, Blanton, Blazek, Bolton, Brinkmann, Brownstein
  et~al.}}]{SDSS-IV_quasars}
\bibinfo{author}{\bibfnamefont{M.}~\bibnamefont{Ata}},
  \bibinfo{author}{\bibfnamefont{F.}~\bibnamefont{Baumgarten}},
  \bibinfo{author}{\bibfnamefont{J.}~\bibnamefont{Bautista}},
  \bibinfo{author}{\bibfnamefont{F.}~\bibnamefont{Beutler}},
  \bibinfo{author}{\bibfnamefont{D.}~\bibnamefont{Bizyaev}},
  \bibinfo{author}{\bibfnamefont{M.~R.} \bibnamefont{Blanton}},
  \bibinfo{author}{\bibfnamefont{J.~A.} \bibnamefont{Blazek}},
  \bibinfo{author}{\bibfnamefont{A.~S.} \bibnamefont{Bolton}},
  \bibinfo{author}{\bibfnamefont{J.}~\bibnamefont{Brinkmann}},
  \bibinfo{author}{\bibfnamefont{J.~R.} \bibnamefont{Brownstein}},
  \bibnamefont{et~al.}, \bibinfo{journal}{Monthly Notices of the Royal
  Astronomical Society} \textbf{\bibinfo{volume}{473}},
  \bibinfo{pages}{4773–4794} (\bibinfo{year}{2017}), ISSN
  \bibinfo{issn}{1365-2966},
  \urlprefix\url{http://dx.doi.org/10.1093/mnras/stx2630}.

\bibitem[{\citenamefont{Bautista et~al.}(2017)\citenamefont{Bautista, Busca,
  Guy, Rich, Blomqvist, du~Mas~des Bourboux, Pieri, Font-Ribera, Bailey,
  Delubac et~al.}}]{SDSS-III_La_forests}
\bibinfo{author}{\bibfnamefont{J.~E.} \bibnamefont{Bautista}},
  \bibinfo{author}{\bibfnamefont{N.~G.} \bibnamefont{Busca}},
  \bibinfo{author}{\bibfnamefont{J.}~\bibnamefont{Guy}},
  \bibinfo{author}{\bibfnamefont{J.}~\bibnamefont{Rich}},
  \bibinfo{author}{\bibfnamefont{M.}~\bibnamefont{Blomqvist}},
  \bibinfo{author}{\bibfnamefont{H.}~\bibnamefont{du~Mas~des Bourboux}},
  \bibinfo{author}{\bibfnamefont{M.~M.} \bibnamefont{Pieri}},
  \bibinfo{author}{\bibfnamefont{A.}~\bibnamefont{Font-Ribera}},
  \bibinfo{author}{\bibfnamefont{S.}~\bibnamefont{Bailey}},
  \bibinfo{author}{\bibfnamefont{T.}~\bibnamefont{Delubac}},
  \bibnamefont{et~al.}, \bibinfo{journal}{Astronomy \& Astrophysics}
  \textbf{\bibinfo{volume}{603}}, \bibinfo{pages}{A12} (\bibinfo{year}{2017}),
  ISSN \bibinfo{issn}{1432-0746},
  \urlprefix\url{http://dx.doi.org/10.1051/0004-6361/201730533}.

\bibitem[{\citenamefont{du~Mas~des Bourboux
  et~al.}(2017)\citenamefont{du~Mas~des Bourboux, Le~Goff, Blomqvist, Busca,
  Guy, Rich, Yèche, Bautista, Burtin, Dawson
  et~al.}}]{La_forests_quasars_cross}
\bibinfo{author}{\bibfnamefont{H.}~\bibnamefont{du~Mas~des Bourboux}},
  \bibinfo{author}{\bibfnamefont{J.-M.} \bibnamefont{Le~Goff}},
  \bibinfo{author}{\bibfnamefont{M.}~\bibnamefont{Blomqvist}},
  \bibinfo{author}{\bibfnamefont{N.~G.} \bibnamefont{Busca}},
  \bibinfo{author}{\bibfnamefont{J.}~\bibnamefont{Guy}},
  \bibinfo{author}{\bibfnamefont{J.}~\bibnamefont{Rich}},
  \bibinfo{author}{\bibfnamefont{C.}~\bibnamefont{Yèche}},
  \bibinfo{author}{\bibfnamefont{J.~E.} \bibnamefont{Bautista}},
  \bibinfo{author}{\bibfnamefont{Ã.}~\bibnamefont{Burtin}},
  \bibinfo{author}{\bibfnamefont{K.~S.} \bibnamefont{Dawson}},
  \bibnamefont{et~al.}, \bibinfo{journal}{Astronomy \& Astrophysics}
  \textbf{\bibinfo{volume}{608}}, \bibinfo{pages}{A130} (\bibinfo{year}{2017}),
  ISSN \bibinfo{issn}{1432-0746},
  \urlprefix\url{http://dx.doi.org/10.1051/0004-6361/201731731}.

\bibitem[{\citenamefont{{Riess} et~al.}(2018)\citenamefont{{Riess},
  {Casertano}, {Yuan}, {Macri}, {Anderson}, {MacKenty}, {Bowers}, {Clubb},
  {Filippenko}, {Jones} et~al.}}]{Riess2018}
\bibinfo{author}{\bibfnamefont{A.~G.} \bibnamefont{{Riess}}},
  \bibinfo{author}{\bibfnamefont{S.}~\bibnamefont{{Casertano}}},
  \bibinfo{author}{\bibfnamefont{W.}~\bibnamefont{{Yuan}}},
  \bibinfo{author}{\bibfnamefont{L.}~\bibnamefont{{Macri}}},
  \bibinfo{author}{\bibfnamefont{J.}~\bibnamefont{{Anderson}}},
  \bibinfo{author}{\bibfnamefont{J.~W.} \bibnamefont{{MacKenty}}},
  \bibinfo{author}{\bibfnamefont{J.~B.} \bibnamefont{{Bowers}}},
  \bibinfo{author}{\bibfnamefont{K.~I.} \bibnamefont{{Clubb}}},
  \bibinfo{author}{\bibfnamefont{A.~V.} \bibnamefont{{Filippenko}}},
  \bibinfo{author}{\bibfnamefont{D.~O.} \bibnamefont{{Jones}}},
  \bibnamefont{et~al.}, \bibinfo{journal}{Astrophysical Journal}
  \textbf{\bibinfo{volume}{855}}, \bibinfo{eid}{136} (\bibinfo{year}{2018}),
  \eprint{1801.01120}.

\end{thebibliography}

\end{document}